\documentclass[usenatbib]{mnras}
\usepackage[utf8]{inputenc}
\usepackage{aas_macros}
\usepackage{amsfonts}
\usepackage{amsmath}
\usepackage{amssymb}
\usepackage{graphicx}
\usepackage[dvipsnames]{xcolor}
\usepackage{hyperref}
\hypersetup{colorlinks=true,allcolors=teal}
\usepackage{microtype}

\newcommand{\Hi}{\textsc{Hi}}

\newcommand{\atot}{\ensuremath{a_{\rm tot}}}
\newcommand{\abary}{\ensuremath{a_{\rm bary}}}

\newcommand{\Msun}{\ensuremath{M_{\odot}}}
\newcommand{\Mh}{\ensuremath{h^{-1}M_{\odot}}}
\newcommand{\Mhsq}{\ensuremath{h^{-2}M_{\odot}}}
\newcommand{\Mpch}{\ensuremath{h^{-1}{\rm Mpc}}}

\newcommand{\msq}{\ensuremath{{\rm \,m\,s}^{-2}}}

\newcommand{\avg}[1]{\ensuremath{\left\langle \,#1\, \right\rangle}}
\newcommand{\e}[1]{\ensuremath{{\rm e}^{#1}}}

\newcommand{\der}{\ensuremath{{\rm d}}}

\newcommand{\eqn}[1]{equation~\eqref{#1}}
\newcommand{\eqns}[1]{equations~\eqref{#1}}

\newcommand{\be}{\begin{equation}}
\newcommand{\ee}{\end{equation}}
\newcommand{\Cal}[1]{\ensuremath{\mathcal{#1}}}

\title[RAR in LCDM]{The radial acceleration relation in a $\Lambda$CDM universe} 
\author[Paranjape \& Sheth]{
Aseem Paranjape$^{1}$\thanks{E-mail: aseem@iucaa.in} \& Ravi K. Sheth$^{2,3}$\thanks{E-mail: shethrk@physics.upenn.edu},
\\  
 $^1$ Inter-University Centre for Astronomy \& Astrophysics,
      Ganeshkhind, Post Bag 4, Pune 411007, India\\
 $^2$ Center for Particle Cosmology, University of Pennsylvania, 209 S. 33rd St., Philadelphia, PA 19104, USA\\
 $^3$ The Abdus Salam International Center for Theoretical Physics, Strada Costiera, 11, Trieste 34151, Italy}

\begin{document}
\label{firstpage}
\pagerange{\pageref{firstpage}--\pageref{lastpage}}
\maketitle

\begin{abstract}
We study the radial acceleration relation (RAR) between the total (\atot) and baryonic (\abary) centripetal acceleration profiles of central galaxies in the cold dark matter (CDM) paradigm. 
We analytically show that the RAR is intimately connected with the physics of the quasi-adiabatic relaxation of dark matter in the presence of baryons in deep potential wells. 
This cleanly demonstrates how the mean RAR and its scatter emerge in the low-acceleration regime ($10^{-12}\msq\lesssim \abary\lesssim10^{-10}\msq$) from an interplay between baryonic feedback processes and the distribution of CDM in dark halos. 
Our framework allows us to go further and study both higher and  lower accelerations in detail, using analytical approximations and a realistic mock catalog of $\sim342,000$ low-redshift central galaxies with $M_r\leq-19$. 
We show that, while the RAR in the baryon-dominated, high-acceleration regime ($\abary\gtrsim10^{-10}\msq$) is very sensitive to details of the relaxation physics,
a simple `baryonification' prescription matching the relaxation results of hydrodynamical CDM simulations is remarkably successful in reproducing the observed RAR \emph{without any tuning}. 
And in the (currently unobserved) ultra-low-acceleration regime ($\abary\lesssim 10^{-12}\msq$), the RAR is sensitive to the abundance of diffuse gas in the halo outskirts, with our default model predicting a distinctive break from a simple power-law-like relation for \Hi-deficient, diffuse gas-rich centrals. 
Our mocks also show that the RAR provides more robust, testable predictions of the $\Lambda$CDM paradigm at galactic scales, with implications for alternative gravity theories, than the baryonic Tully-Fisher relation. 
\end{abstract}


\section{Introduction}
\label{sec:intro}
\noindent
Gravitational interactions at  galactic  scales offer a fertile testing ground for competing theories of gravitation. The highly successful Lambda-cold dark matter ($\Lambda$CDM) paradigm attributes all gravitational interactions at these scales to the Newtonian limit of general relativity, but postulates the existence of a collisionless (or dark) matter component that pervades the cosmos \cite[for a recent review, see][]{salucciAApR}. In stark contrast, alternative proposals such as Modified Newtonian Dynamics \citep[MOND,][]{milgrom83} attempt to explain extra-Galactic observations, particularly galactic rotation curves, using Standard Model physics alone (i.e., without a dark component), but alter the nature of gravity at these scales. MOND, in particular, postulates a new, fundamental acceleration scale $a_0\sim10^{-10}\msq$ to  segregate the high-acceleration regime of Newtonian dynamics from the low-acceleration regime where the nature of gravity is modified. MOND is just one of a growing number of modified gravity models \cite[for a recent review, see][]{bertoneTait18}.

Observationally, such competing ideas are potentially amenable to testing using empirical correlations between the dynamical, gravitating mass of a system and the light we observe from it. Among the several such mass-to-light scalings that are known to exist for galaxies of different types \citep{fj76,tf77,msbdb00},
the `radial acceleration relation' \citep[RAR,][]{mls16} has recently emerged as an intriguing new potential test of gravity.

The RAR is usually expressed as the relation between the centripetal acceleration profile $\atot(r)$ due to all gravitating components (in $\Lambda$CDM, these would be baryonic and dark matter), and the Newtonian contribution $\abary(r)$ to this profile from the baryonic components alone. In terms of the galactic rotation curve $v_{\rm rot}(r)$ and its baryonic contribution $v_{\rm bary}(r)$ (these will be defined below), we have 
\be
\atot(r) = v_{\rm rot}^2(r) / r\,,
\label{eq:atot-def}
\ee
and
\begin{align}
\abary(r) &= v_{\rm bary}^2(r) / r\,.
\label{eq:abary-def}
\end{align}
The RAR and its close cousin, the baryonic Tully-Fisher relation \citep[BTFR,][]{msbdb00}, have been extensively discussed in the literature, especially in the context of MOND versus $\Lambda$CDM \citep[see, e.g., the review by][see also below]{mcgaugh15} 
In the $\Lambda$CDM framework, unlike MOND, there is no fundamental acceleration scale. Correlations such as the RAR and BTFR, to the extent that they are predicted by $\Lambda$CDM, are necessarily emergent phenomena that result from a complex combination of many underlying correlations. The fact that the \emph{observed} BTFR and especially the RAR have low scatter, makes it very interesting to ask how the emergence of these relations in $\Lambda$CDM fares against observations \citep[see, e.g.,][for a discussion of the constraints on physical models of the Tully-Fisher relation]{courteau+07}. Several studies have followed this line of reasoning and used hydrodynamical CDM simulations of, both, small samples of objects as well as cosmological volumes, to quantify the BTFR and RAR expected in $\Lambda$CDM \citep[e.g.,][]{sg16,sales+17,kw17,ludlow+17,tenneti+18,grpp18}. 

Focusing on the RAR (we discuss the BTFR separately later), a general trend is that most hydrodymanical CDM simulations that broadly reproduce observed galaxy properties do, in fact, also naturally produce a tight RAR (e.g., \citealp{kw17}, although see \citealp{milgrom16}). However, the details of the median trend and the scatter around it do not always agree with the observed ones \citep[e.g.,][]{ludlow+17,tenneti+18}, and it is usually difficult to assess whether the differences are fundamental (e.g., due to specifics of baryonic feedback physics) or caused by widely different sample definitions and other technical choices in measuring rotation curves. For example, the EAGLE simulations produce an RAR similar to the observed one but with an inferred acceleration scale $a_0$ higher by about a factor 2 \citep{ludlow+17}, while the RAR in the MassiveBlack-II simulation is closer to a  power law with no intrinsic acceleration scale \citep{tenneti+18}.

Several authors have attempted to build an analytical understanding of the RAR in a $\Lambda$CDM universe \citep[see][for early work]{vdbd00}. \citet{whd19} have argued that the RAR is a simple algebraic outcome of the BTFR, although they do not address the emergence of the BTFR itself. \citet{gbfh20} have attempted to explain the emergence of a characteristic acceleration scale from the physics of stellar feedback, expressing $a_0$ using fundamental constants. 
The emergence of the RAR and related scalings in $\Lambda$CDM is, in general, easier to appreciate using empirical models to connect dark matter to baryons, along with (semi-)analytical modelling for producing rotation curves. This approach has been adopted by several authors recently using the  subhalo abundance matching (SHAM) technique \citep[e.g.,][]{dw15,desmond17,navarro+17}. A common thread in these studies is that the $\Lambda$CDM RAR is a complicated but natural outcome of a combination of the SHAM association of stellar mass to dark halos, the requirement that galaxy disk sizes obey the observationally constrained scaling with halo properties, and the magnitude of the `backreaction' of the baryonic material on the dark matter profile in the inner halo.

In this work, we present new analytical insights into the structure of the RAR, and the underlying physics that determines this structure, in the $\Lambda$CDM paradigm. Specifically, we show that the physics of \emph{quasi-adiabatic relaxation} of the dark matter profile in the presence of baryons, particularly in the inner, baryon-dominated regions of the halo, \emph{plays a key role in establishing both the median and scatter of the RAR} for any galaxy sample. 
Although previous work \citep[e.g.,][]{desmond17} has noticed the relevance of this relaxation physics to the RAR, its full impact on the RAR has not been appreciated to date \citep[e.g.,][discuss the RAR in the \emph{absence} of any baryonic effect on the dark matter]{navarro+17}. We believe this is largely due to the common practice of expressing the RAR as the functional dependence of \atot\ on \abary\ \citep[e.g.,][]{mls16,lmsp17,kw17,ludlow+17,navarro+17,desmond17,tenneti+18,dpsf19,tian+20}, which can easily mask small but significant differences between alternative physical models in the predicted \emph{approach} of $\atot\to\abary$ at large \abary. As argued by \citet{cbsg19}, the baryon-dominated, high-acceleration regime  ($\abary\gtrsim10^{-10}\msq$) of the RAR is better probed by expressing the quantity
\be
\Delta_a \equiv \atot/\abary-1\,,
\label{eq:Delta_a-def}
\ee
as a function of $\abary$. In the language of \citet{mcgaugh99}, $\Delta_a$ can be thought of as a `residual mass discrepancy'. We exclusively use this formulation of the RAR in the present work.

We augment our analytical calculations with  measurements of the RAR in a mock galaxy catalog containing a cosmologically representative sample of central galaxies with realistic baryonic properties, including stellar mass and cold as well as hot gas, along with their spatial distributions. This mock is based on the algorithm recently presented by \citet{pcs21} and is described below. The use of mock galaxies with numerically sampled rotation curves allows us to extensively explore the sensitivity of the RAR to changes not only in the underlying physics and baryon-dark matter scalings, but also to effects of sample selection and other technical aspects of rotation curve estimation. Our primary goal is to emphasize and disentangle conceptual issues, rather than perform a detailed comparison with observations.
We therefore ignore observational errors and focus on the intrinsic predictions that follow from our analytical arguments and mock catalogs. As such, we deal only with `perfectly measured' rotation curves in this work \citep[see][for more careful comparisons with observed data sets]{desmond17}.

The paper is organized as follows. 
In section~\ref{sec:mocks}, we briefly describe the numerical algorithm and $N$-body simulation box underlying the mock galaxy catalog we use in this work. 
In section~\ref{sec:physics:analytical}, we present analytical calculations that show how any prescription for quasi-adiabatic relaxation and the associated baryon-dark matter scalings (section~\ref{subsec:rdm}) leads directly to a prediction for the RAR of each individual galaxy, and hence of any population of galaxies (section~\ref{subsec:xiRAR}). 
Appendix~\ref{app:analytic} builds on these analytical results to construct an approximate but fully analytic RAR which allows us to predict the shape and tightness of the RAR in various limits.
In section~\ref{sec:physics:mocks}, we explore the RAR of our mock galaxies for various choices of relaxation physics prescription, sample selection, baryon-dark matter scaling, and technical details such as rotation curve sampling. This exercise allows us to put all our analytical arguments to the test.  
In section~\ref{sec:btfr}, we discuss in detail the predictions of our mocks for the BTFR, highlighting the pitfalls of over-interpreting BTFR measurements which, unlike the RAR, are inherently unstable to variations in technical details of the analysis.
We conclude in section~\ref{sec:conclude}. 

Throughout, $m_{\rm vir}$ and $R_{\rm vir}$ refer to the total halo mass and virial radius. In keeping with the literature on quasi-adiabatic relaxation, on which we rely heavily, we define $R_{\rm vir}\equiv R_{\rm 200c}$, the radius at which the enclosed halo-centric density becomes 200 times the critical density $\rho_{\rm crit}$ of the Universe, so that $m_{\rm vir} = (4\pi/3)R_{\rm vir}^3\times200\rho_{\rm crit}$. All our results assume a spatially flat $\Lambda$CDM background cosmology, with parameters $\{\Omega_{\rm m},\Omega_{\rm b},h,n_{\rm s},\sigma_8\}$ given by $\{$0.276, 0.045, 0.7, 0.961, 0.811$\}$, compatible with the 7-year results of the \emph{Wilkinson Microwave Anisotropy Probe} experiment \citep[WMAP7,][]{Komatsu2010}. We will denote the base-10 (natural) logarithm as log (ln). 

\section{Mock catalogs}
\label{sec:mocks}
Our results are based on a mock galaxy catalog constructed using the algorithm described in  detail by \citet[][hereafter, PCS21]{pcs21}. Below, we briefly summarise this algorithm and the $N$-body simulation that is populated with mock galaxies, followed by a discussion of the baryonic components and associated rotation curve of each mock central galaxy.

\subsection{Simulation and mock algorithm}
\label{subsec:sims-algo}
We use one realisation of the ${\rm L}300\_{\rm N}1024$ simulation configurations discussed by PCS21. This is a gravity-only simulation with $1024^3$ particles in a $(300\Mpch)^3$ cubic box, performed using the code \textsc{gadget-2} \citep{springel:2005}\footnote{\url{http://www.mpa-garching.mpg.de/gadget/}} with halos identified using the code \textsc{rockstar} \citep{behroozi13-rockstar}.\footnote{\url{https://bitbucket.org/gfcstanford/rockstar}} Further details of the simulation can be found in \citet{pa20}. 

The PCS21 algorithm, which is based on the halo occupation distribution (HOD) models  calibrated by \citet{pcp18} and \citet{ppp19}, populates host halos in this box with mock central and satellite galaxies, producing a luminosity-complete sample of galaxies with an $r$-band absolute magnitude threshold $M_r\leq-19$. 
In addition to the $r$-band magnitude, each mock galaxy is assigned realistic values of $g-r$ and $u-r$ colours and stellar mass $m_\ast$. A fraction of these galaxies is also  assigned non-zero values of neutral hydrogen (\Hi) mass $m_{\Hi}$.
The HOD models underlying this algorithm are constrained by the observed abundances and clustering of optically selected galaxies in the Sloan Digital Sky Survey \citep[SDSS,][]{york+00},\footnote{\url{www.sdss.org}} and of \Hi-selected galaxies in the ALFALFA survey \citep{giovanelli+05}. PCS21 presented extensive tests of the algorithm, along with a detailed discussion of cross-correlation statistics between optical and \Hi-selected samples that are predicted by the  algorithm.

In this work, we focus only on central galaxies, whose host halos are `baryonified' by the PCS21 algorithm as discussed below.  The ${\rm L}300\_{\rm N}1024$ box described above contains approximately $342,000$ central galaxies with $M_r\leq-19$. The median along with 16th and 84th percentiles of $\log[m_{\rm vir}/(\Mh)]$ for the host halos of these centrals is $11.64^{+0.54}_{-0.33}$. At fixed mass, halo concentrations have a mass-independent Lognormal scatter of $\sigma_{\ln c_{\rm vir}}=0.16\ln(10)$. For the overall distribution of central galaxy hosts, this gives a median with 16th and 84th percentiles of $c_{\rm vir} = 7.7^{+3.8}_{-2.6}$. Here $c_{\rm vir}=R_{\rm vir}/r_{\rm s}$, with $r_s$ the scale radius of the halo returned by \textsc{rockstar} by fitting a \citet*[][NFW]{nfw96} profile.

\subsection{Baryonification scheme}
\label{subsec:baryonify}
The PCS21 algorithm uses a modified version of the baryonification prescription of \citet[][hereafter, ST15]{st15} to model the spatial distributions of a number of baryonic components in each central galaxy and its  host halo. These include: 
\begin{itemize}
    \item A spherical distribution of stars in the central galaxy (`cgal') with half-light radius $R_{\rm hl}$ whose relation with the halo radius $R_{\rm vir}$ is constrained by observations \citep{kravtsov13}. The corresponding mass fraction is $f_{\rm cgal}=m_\ast/m_{\rm vir}$. In principle, we could also model the stellar distribution as a combination of a disk and a bulge, which  we leave for future work.
    \item A 2-dimensional axisymmetric \Hi\ disk (`\Hi') with scale length $h_{\Hi}$, for centrals with $m_{\Hi}>0$, with the $h_{\Hi}$-$m_{\Hi}$ relation being constrained by observations \citep[][see equation~8 of PCS21]{wang+16-HI}. The corresponding mass fraction is $f_{\Hi}=1.33\,m_{\Hi}/m_{\rm vir}$, with the prefactor accounting for Helium correction. (The \Hi\ disk was not modelled by ST15.)
    \item A spherical distribution of bound hot gas (`bgas') in hydrostatic equilibrium. The halo mass dependence of the corresponding mass fraction $f_{\rm bgas}$ is constrained by X-ray cluster observations at $m_{\rm vir}\gtrsim10^{13}\Mh$ using a 2-parameter model and extrapolated to lower masses where needed. We will discuss the sensitivity of our results to these parameter values later.
    \item Expelled gas (`egas') or the circum-galactic medium (CGM). As discussed by PCS21, for rotation curve  modelling this is essentially a uniform density distribution inside $R_{\rm vir}$, so that the specific value of the free parameter used by ST15 to model its distribution does not affect any of the analysis below. The corresponding mass fraction $f_{\rm egas}$ is constrained by baryonic mass conservation by demanding\footnote{This is violated by a small fraction ($\sim1\%$) of objects with $M_r\leq-19$ for which the sum $f_{\rm cgal} + f_{\Hi} + f_{\rm bgas}$ exceeds $\Omega_{\rm b}/\Omega_{\rm m}$ (which in turn are dominated by objects having $f_{\rm cgal} + f_{\Hi} > \Omega_{\rm b}/\Omega_{\rm m}$). For such objects, we follow PCS21 and set $f_{\rm egas}=0$ without changing any of the other baryonic mass fractions, so that $f_{\rm bary} >\Omega_{\rm b}/\Omega_{\rm m}$. Overall mass conservation then implies that the corresponding dark matter fraction $f_{\rm rdm}=1-f_{\rm bary}$ (see section~\ref{sec:physics:analytical}) is smaller than $1-\Omega_{\rm b}/\Omega_{\rm m}$ for these objects. \label{fn:masscons}}
    \be
    f_{\rm bary} \equiv f_{\rm cgal} + f_{\Hi} + f_{\rm bgas} + f_{\rm egas} = \Omega_{\rm b}/\Omega_{\rm m} \simeq 0.163\,.
    \label{eq:baryonicmasscons}
    \ee
\end{itemize}
In addition to modelling the \Hi\ disk, the PCS21 version of baryonification also departs from ST15 by truncating and normalising all mass profiles at the halo virial radius rather than at infinity. As discussed by \citet{arico+20}, this considerably simplifies the implementation of this scheme while still maintaining its accuracy in our regime of interest. Further details of the numerical implementation, as well as all the underlying scalings of baryonic mass fractions and galaxy sizes with halo properties, can be found in section~3.2 of PCS21.  
Baryonification schemes of this type have been shown to successfully reproduce the small-scale matter power spectrum and bispectrum of cosmological hydrodynamical simulations \citep[e.g.,][]{chisari+18,arico+21}.

The rotation curve $v_{\rm rot}(r)$ for each mock galaxy is calculated using equation~(11) of PCS21, which can be rewritten as
\begin{align}
v_{\rm rot}^2(r) &= v_{\Hi}^2(r) + \sum_{\alpha} \frac{Gm_{\alpha}(<r)}{r} +  \frac{Gm_{\rm rdm}(<r)}{r}\notag\\
&\equiv v_{\rm bary}^2(r) + \frac{Gm_{\rm rdm}(<r)}{r}\,,
\label{eq:vrot-def}
\end{align}
where, in the first line, $v_{\Hi}^2(r)$ is the \Hi\ disk contribution (equation 10 of PCS21), the sum runs over $\alpha\in\{{\rm bgas},{\rm cgal},{\rm egas}\}$, $m_{\alpha}(<r)$ is the mass of component $\alpha$ enclosed in radius $r$ and $m_{\rm rdm}(<r)$ is the corresponding mass of the `relaxed' dark matter component which we discuss in detail in the next section, and the second line defines the baryonic contribution $v_{\rm bary}^2(r)$.

Below, we will also use the total (sphericalised) mass profile contained in radius $r$, which can be split into contributions from baryons and the relaxed dark matter component,
\begin{align}
m_{\rm tot}(<r) &= m_{\rm bary}(<r) + m_{\rm rdm}(<r)\notag\\
&= \sum_\chi\,m_\chi(<r) + m_{\rm rdm}(<r)\,,
\label{eq:mtot-def}
\end{align}
where the sum in the second line runs over $\chi\in\{\textrm{bgas,\,cgal,\,egas,\,\Hi}\}$. 

For later use, we also calculate an integrated baryonic mass $M_{\rm bary}$ \citep[e.g.,][]{lmsp17,sales+17} for each  central as the sum of the masses of stars and cold gas contained inside the radius $r=2R_{\rm h,bary}$, where $R_{\rm h,bary}$ is the radius which encloses half the mass of stars and cold gas\footnote{In practice, we determine $2R_{\rm h,bary}$ by sampling the rotation curve using $200$ logarithmically spaced points in the range $(0.001,1)\times R_{\rm vir}$ for each central galaxy.}:
\be
M_{\rm bary} = m_{\rm cgal}(<2R_{\rm h,bary}) + m_{\Hi}(<2R_{\rm h,bary})\,,
\label{eq:Mbary-def}
\ee
and where $m_{\Hi}(<r)$ includes the Helium correction mentioned above, so that $m_{\Hi}(<R_{\rm vir})=1.33\,m_{\Hi}$. Our use of a 3-dimensional half-mass radius to define $M_{\rm bary}$ can, in principle, lead to systematic effects when comparing with observations which typically use projected sizes for measuring $M_{\rm bary}$. For such analyses below, we have checked that replacing $M_{\rm bary}$ with the total $m_\ast+1.33m_{\Hi}$ for each galaxy leads to identical conclusions, i.e., our results are expected to be insensitive to the exact definition of $M_{\rm bary}$.

As discussed in the Introduction, the radial acceleration relation is then the dependence of $\Delta_a=\atot/\abary-1$  on \abary, with \atot\ and \abary\ given by \eqns{eq:atot-def} and~\eqref{eq:abary-def}, respectively. 
Notice that $v_{\rm bary}$, and hence $\abary$, contains contributions from both spherical as well as axisymmetric components.  This is consistent with observational analyses of the RAR \citep[see, e.g.,][]{mls16}.

\section{Physics of the RAR: Analytical insights}
\label{sec:physics:analytical}
\noindent
Thus far, we have not commented on the shape of the relaxed dark matter profile $m_{\rm rdm}(<r)$. As we discuss in this section, this is a key component in determining the shape of the mean RAR.

\begin{figure*}
    \centering
    \includegraphics[width=0.85\textwidth,trim=8 5 5 5,clip]{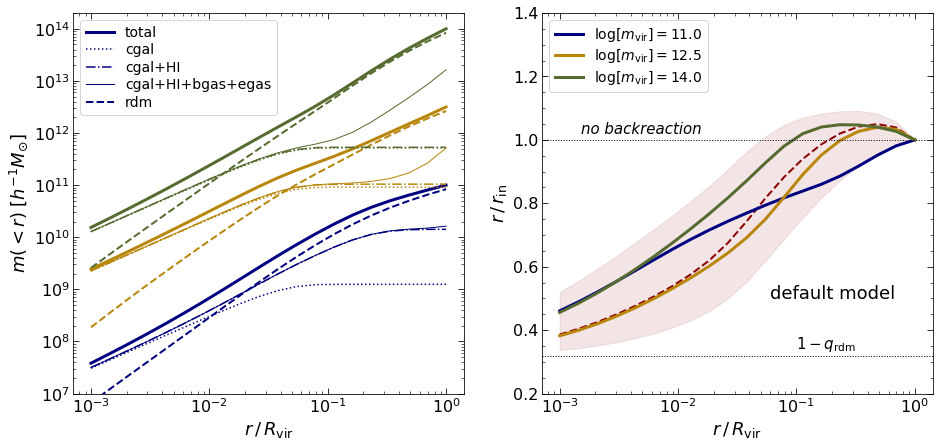}
    \caption{{\bf Relaxation physics}. \emph{(Left panel:)} Mass profiles for three examples of baryonified halos hosting an NGC99-like galaxy (curves with different colours), with different combinations of components shown using the linestyles indicated in the legend. Halo concentrations, stellar masses and \Hi\ disk sizes were set using scaling relations from the literature \citep[see][PCS21]{pcs21}, while the \Hi\ mass was fixed to $m_{\Hi}=10^{9.83}\Mhsq$ in each case. Other baryonic fractions were set as described in the text. The two lower mass halos are the same as shown in figure 4 of PCS21. \emph{(Right panel:)} Relaxation ratio $\xi = r/r_{\rm in}$ for the dark matter profile computed using \eqns{eq:xi-soln} and \eqref{eq:X-def} with $q_{\rm rdm}=0.68$. The thick solid curves show $\xi$ for the three halos from the left panel. The dashed red curve and band respectively show the median and central $95\%$ range of $\xi$ for the entire luminosity-complete mock catalog used in the text. The lower horizontal line shows the theoretical lower bound of $1-q_{\rm rdm}$ (see text). The upper horizontal line indicates unity, the solution when baryons do not affect the  dark matter profile. Values of $\xi$ less (greater)  than unity correspond to contraction (expansion) of the dark matter profile due to the presence of baryons.}
    \label{fig:relaxation-ratio}
\end{figure*}

\subsection{Quasi-adiabatic relaxation}
\label{subsec:rdm}
\noindent
In the default PCS21 model, $m_{\rm rdm}(<r)$ is calculated assuming complete spherical symmetry  for all components, and assuming that the dark matter quasi-adiabatically relaxes (approximately conserving angular momentum) in response to the baryonic components. The details of the procedure can be found in ST15 or Appendix~A of PCS21 and are briefly summarised below. This relaxation can be described using a function $\xi(r)$ defined as
\be
\xi \equiv r/r_{\rm in}\,,
\label{eq:xi-def}
\ee
where $r_{\rm in}$ is the initial radius of a spherical dark matter element which eventually relaxes to a final radius $r$. The equation governing the form of $\xi$ can be written in general as
\be
\xi = 1 + \Cal{X}\left(\frac{m_{\rm udm}(<r_{\rm in})}{m_{\rm tot}(<r)}\right)\,,
\label{eq:xi-soln}
\ee
where $m_{\rm udm}(<r_{\rm in})$ is the unrelaxed dark matter profile.
We approximate this using the NFW form in this work (although see below).
The function $\Cal{X}(y)$ in the ST15 model, which was adopted by PCS21, is given by
\be
\Cal{X}(y) = q_{\rm rdm}\,(y-1)\,.
\label{eq:X-def}
\ee
Here $q_{\rm rdm}$ is a parameter controlling the level of angular momentum conservation, with $q_{\rm rdm} = 1$ for perfect conservation and $q_{\rm rdm}=0$ for no baryonic backreaction. The default model from PCS21 follows the ST15 prescription and sets $q_{\rm rdm}=0.68$. Equation~\eqref{eq:xi-soln} is then solved iteratively to obtain $\xi(r)$, using which the relaxed dark matter profile satisfies (see Appendix~A of PCS21)
\be
m_{\rm rdm}(<r) = f_{\rm rdm}\,m_{\rm udm}(<r/\xi)\,,
\label{eq:mrdm}
\ee
where $f_{\rm rdm}$ is the mass fraction of dark matter inside the host halo's virial radius; due to \eqn{eq:baryonicmasscons}, this is set to $f_{\rm rdm}=1-\Omega_{\rm b}/\Omega_{\rm m}$ in this work for all but the small fraction of objects discussed in footnote~\ref{fn:masscons}.

Figure~\ref{fig:relaxation-ratio} shows the numerically computed relaxation ratio $\xi$ in the ST15 model \emph{(right panel)} for three examples of baryonified halos whose mass profiles are shown in the \emph{left panel}. The \emph{right panel} shows that there is a lower limit to $\xi$ because $y\geq 0$ in \eqn{eq:X-def} (being the ratio of masses, $y$ cannot be negative at any $r$).  Moreover, while the ST15 model leads to a contraction of the dark matter profile throughout the least massive halo, it predicts an expansion in the outskirts of more massive halos. A comparison with the \emph{left panel} shows that this happens in regions where the fraction of bound and/or expelled gas is higher than that of stars and the \Hi\ disk (compare the thin solid lines which show all baryons with the dash-dotted lines showing only the stellar and \Hi\ component). The dashed red curve and band in the \emph{right panel} respectively show the median and central $95\%$ range of $\xi$ for the entire mock catalog used below.

Strictly speaking, the assumption of perfect spherical symmetry is not valid due to the presence of the axisymmetric baryonic disk, as well as the fact that dark matter halos in gravity-only simulations are triaxial in general. Including these non-spherical effects analytically and calculating a triaxial $\vec{\xi}(\vec{r})$ is quite difficult. Interestingly, though, the results of hydrodynamical simulations show that baryonic backreaction actually tends to make the dark matter distribution after relaxation \emph{more} spherical \citep{dubinski94,kazantzidis+04,abadi+10,cpta21}. We therefore expect that, in practice, our spherical assumption will lead to an accurate average description of quasi-adiabatic relaxation. We intend to explore the effects of asphericity in the relaxation process, along with detailed comparisons to hydrodynamical simulations, in future work.

\subsection{Relaxation ratio and the RAR}
\label{subsec:xiRAR}
\noindent
Equation~\eqref{eq:xi-soln} allows us to appreciate an intimate connection between the level of angular momentum conservation and the shape of the RAR. For the spherically symmetric case assumed above,  \eqns{eq:xi-def} and~\eqref{eq:mrdm} give the identity
\be
\frac{m_{\rm udm}(<r_{\rm in})}{m_{\rm tot}(<r)}  = \frac{1}{f_{\rm rdm}}\,\frac{\atot(r)-\abary(r)}{\atot(r)}\,.
\label{eq:massratio}
\ee
Using this, \eqn{eq:xi-soln} can be formally inverted and, after some straightforward algebra, brought to the form
\be
 \Delta_a  = \frac{\Lambda}{1-\Lambda}\,,
\label{eq:ardmByabary}
\ee
where $\Delta_a$ was defined in \eqn{eq:Delta_a-def}, and
\be
\Lambda \equiv f_{\rm rdm}\,\Cal{X}^{-1}(\xi-1)\,,
\label{eq:Lambda-def}
\ee
with $\Cal{X}^{-1}(z)=y$ being the inverse function of $\Cal{X}(y)=z$.

Equation~\eqref{eq:ardmByabary} is remarkable because it shows that, as a function of the relaxation ratio $\xi$, the RAR has $\emph{zero scatter}$ in the spherical baryonification model, \emph{regardless} of the exact functional form of $\Cal{X}(y)$ which sets the mean relation.  
For our default choice of $f_{\rm rdm}=1-\Omega_{\rm b}/\Omega_{\rm m}$, it is clear from \eqn{eq:Lambda-def} that \emph{the scatter in the RAR as defined in the literature arises solely from the scatter between $\xi$ and $\abary$.} 
If we think of these functions as $\xi = \xi(r|{\rm bary},{\rm dm})$ and $\abary = \abary(r|{\rm bary})$, then, at fixed $r$ and for a given baryonic configuration, this scatter is caused predominantly by the object-to-object variation in halo mass and concentration for different central galaxies with this baryonic configuration. There could be some additional scatter at fixed halo mass and concentration if, for example, multiple baryonic configurations happen to lead to the same value of $\xi$ but different $\abary$, or vice-versa.  This is, of course, very different from MOND which predicts that the RAR should have \emph{no} intrinsic scatter.

For the specific choice of $\Cal{X}$ in \eqn{eq:X-def} adopted in this work, \eqn{eq:ardmByabary} simplifies to
\be
 \Delta_a = \frac{(\xi-1) + q_{\rm rdm}}{q_{\rm rdm}/f_{\rm rdm} - (\xi-1) - q_{\rm rdm}}\,.
\label{eq:ardmByabary-thiswork}
\ee
To glean some analytical insights into the implications of \eqn{eq:ardmByabary} or \eqn{eq:ardmByabary-thiswork}, it is useful to analyse the result perturbatively in the case $0<q_{\rm rdm}\ll1$, i.e., in the limit of small baryonic backreaction. This is the same limit as studied by \citet{navarro+17}, who ignored baryonic backreaction and focused on explaining the origin of the RAR in the low-acceleration regime using various baryon-dark matter scalings. At lowest order in $q_{\rm rdm}$, this leads to 
\begin{align}
\xi-1 \simeq q_{\rm rdm}\,\left(\frac{f_{\rm bary}\,m_{\rm udm}(<r) - m_{\rm bary}(<r)}{f_{\rm rdm}\,m_{\rm udm}(<r) + m_{\rm bary}(<r)}\right)\,.
\label{eq:(xi-1):pert}
\end{align}
Plugging this into \eqn{eq:ardmByabary-thiswork} gives, after some simplification,
\be
\Delta_a \simeq \frac{f_{\rm rdm}\,m_{\rm udm}(<r)}{m_{\rm bary}(<r)}\,,
\label{eq:ardmByabary-pert}
\ee
which is an eminently sensible result. This also shows that the RAR in the limit of no baryonic backreaction can be expected to have a large scatter as a function of $\abary$, since the dark matter profile $m_{\rm udm}(<r)$ in the numerator of \eqn{eq:ardmByabary-pert} is  decoupled from $\abary(r)\sim m_{\rm bary}(<r)/r^2$, apart from the baryon-dark matter scalings that relate halo mass and concentration to baryonic mass fractions and sizes.  
Appendix~\ref{app:analytic} shows that, if the initial mass distribution is similar to an NFW profile, then the RAR is amenable to analytic treatment, even when backreaction is large. In particular, one can analytically estimate the RAR of individual galaxies such as the ones depicted by the thick solid lines in figure~\ref{fig:relaxation-ratio}. The resulting dependence of the median and scatter of the RAR on various halo and galaxy properties then provides an analytic understanding of the trends we discuss below using numerically sampled mock galaxies.

\section{Results from mocks}
\label{sec:physics:mocks}
\noindent
With these analytical arguments in hand, we now explore the RAR in the mock catalogs described in section~\ref{sec:mocks} by varying the underlying baryonification choices of the PCS21 algorithm, as well as selecting galaxy samples using various criteria.

\subsection{Default model}
\label{subsec:default}
\begin{figure}
    \centering
    \includegraphics[width=0.425\textwidth,trim=5 10 5 5,clip]{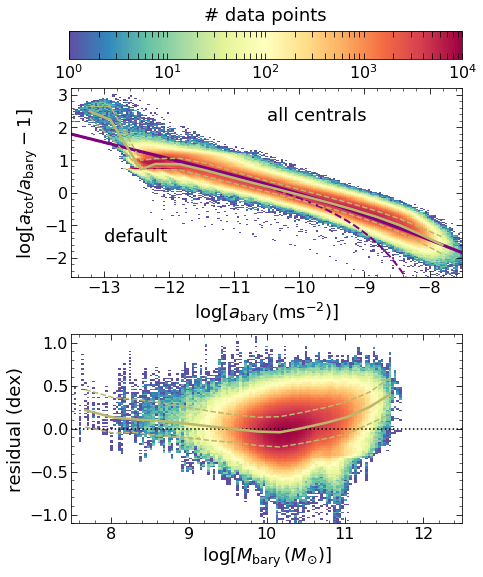}
    \caption{{\bf Default RAR.} \emph{(Top panel:)} The RAR ($\Delta_a$ from equation~\ref{eq:Delta_a-def} as a function of \abary) defined by central galaxies in one mock with our default baryonification model having $q_{\rm rdm}=0.68$ (see sections~\ref{subsec:baryonify} and~\ref{subsec:rdm}). The coloured histogram counts measurements from all centrals, with rotation curves sampled at $20$ logarithmically spaced points in the range $(0.001,1)\times R_{\rm vir}$ for each object. 
    Yellow solid and dashed lines show the median and central $68\%$ region of the distribution in bins of \abary.
    Solid and dashed purple curves show $\Cal{F}(\abary/a_0)-1$ using equations~(\ref{eq:RAR-simple}) and~(\ref{eq:RAR-mcgaugh}), respectively, with parameters as described in the text. 
    The measured median relation agrees well with the solid purple curve at low and high \abary.
    The break in the median relation at ultra-low accelerations ($\abary\lesssim10^{-12}\msq$) is discussed in detail in the text.
    \emph{(Bottom panel:)} Residuals relative to the solid purple curve, computed as $\log[\Delta_a/(\Cal{F}(\abary/a_0)-1)]$, using $\Cal{F}(x)$ from \eqn{eq:RAR-simple}, for each point $(\abary,\atot)$ having $\abary>10^{-12}\msq$, shown as a function of $M_{\rm bary}$ which is assigned for each galaxy using \eqn{eq:Mbary-def} (so each galaxy contributes a vertical streak). Yellow curves now show the median and central $68\%$ of the distribution of the residuals in bins of $M_{\rm bary}$.}
    \label{fig:rar-fullsample}
\end{figure}

The coloured histogram in the \emph{top panel} of figure~\ref{fig:rar-fullsample} shows the RAR -- the horizontal axis shows $\log[\abary(\msq)]$ and the vertical axis shows $\log[\Delta_a]$ -- of the full sample of central galaxies with $M_r\leq-19$ in one mock ($\sim342,000$ objects) for our default baryonification model.  We calculated $\atot(r)$ and $\abary(r)$ on $20$ logarithmically spaced values of $r$ in the range $(0.001,1)\times R_{\rm vir}$ for each central galaxy (we explore the effects of changing this sampling choice below). Our results therefore explore not only the inner, baryon-dominated parts of each halo, but also the halo outskirts corresponding to  the ultra-low-acceleration regime ($\abary\lesssim10^{-12}\msq$) which is as yet observationally unconstrained.

The dashed and solid purple curves show $\log[\Cal{F}(\abary/a_0)-1]$ with $\Cal{F}(x)$ being, respectively, the MOND-inspired calibration for $\atot/\abary$ from equation~(4) of \citet{mls16}, 
\be
 \Cal{F}(x) = \frac{1}{1-\e{-\sqrt{x}}}\,,
\label{eq:RAR-mcgaugh}
\ee
and equation~(4) of \citet{cbsg19}, 
\be
 \Cal{F}(x) = \left[\,\frac12 + \sqrt{\frac14+\frac{1}{x^\nu}}\,\right]^{1/\nu}\,,
\label{eq:RAR-simple}
\ee
with $a_0=1.2\times10^{-10}\msq$ (\citealp{mls16} denote this as $g_\dagger$).  
For $x\ll 1$ both functions scale as $\Cal{F}\to 1/\sqrt{x}$, and both tend to unity when $x\gg 1$ (which simply reflects the fact that this is the limit in which baryons dominate).  However, the approach to this baryon-dominated limit is different:  
$\Delta_a = \Cal{F} - 1\to\e{-\sqrt{x}}$ for \eqn{eq:RAR-mcgaugh}   whereas $\Delta_a \to x^{-\nu}/\nu$ for \eqn{eq:RAR-simple}.  
The solid curve shows $\nu=0.8$, which \citet{cbsg19} argue fits the observed RAR well, particularly at $\abary\ge a_0$ where ellipticals dominate. However, $\nu=1$, which lies approximately midway between the solid and dashed curves, may provide a better description of the RAR defined by spirals \citep[][see also Appendix~\ref{app:hernquist-rar}]{famaeyBinney05, sn07, efe20}.

The solid yellow curve shows the median $\avg{\Delta_{a,{\rm mock}}}$ of the distribution in bins of $\abary$, while the dashed yellow curves show the corresponding 16th and 84th percentiles.\footnote{To calculate the yellow curves, we use 17 linearly spaced bins in $\log[\abary/(\msq)]$ in the range $(-13.5,-7.5)$, discarding bins containing fewer than $10$ data points. The location of the curves on the horizontal axis is taken to be the  median \abary\ of each bin.} We see that the median RAR of our default mock is in remarkably good agreement with the solid purple curve for $\abary\gtrsim10^{-12}\msq$, i.e., throughout the low- and high-acceleration regimes. (Quantitatively, $|\avg{\Delta_{a,{\rm mock}}}/\Delta_{a,{\rm eqn\,18}}-1|\lesssim20\%$ in this range.) Since the solid purple curve was shown by \citet{cbsg19} to be a good description of the observed RAR, \emph{this is a non-trivial success of our default model, with no additional tuning} beyond what was already discussed by PCS21 to match other observations. 
The scatter around the median relation is typically $\sigma_{\log[\atot]}\sim0.075$ dex for $\abary>10^{-12}\msq$. (Note that the scatter in $\log[\Delta_a]$ seen in the figure is considerably larger; we report the scatter in $\log[\atot]$ in the text for ease of comparison with the literature.)
The RAR in the (as yet unobserved) ultra-low acceleration regime of $\abary\lesssim10^{-12}\msq$ sharply breaks away from the extrapolation of \eqns{eq:RAR-simple} and~\eqref{eq:RAR-mcgaugh}, first dipping below at $\abary\simeq10^{-12}\msq$ and then rising steeply at $\abary\lesssim10^{-12.4}\msq$. We have found that the cloud with very few points at the top left of the distribution ($\log[\abary]\lesssim-12.6,\log[\Delta_a]\gtrsim1.4$) is dominated by objects having extremely low $m_\ast$ ($\sim10^6\Mhsq$), which are likely numerical artefacts in the statistical sampling of the colour-dependent mass-to-light ratio in the PCS21 algorithm. We will therefore ignore the regime $\abary\lesssim10^{-12.6}\msq$ in the discussion below. We will, however, later explore the nature of the galaxies which lead to the dip and rise near $\abary\simeq10^{-12}\msq$. For now, we simply note that our results constitute predictions for the ultra-low-acceleration regime \citep[see also][]{obln20}.\footnote{Recently, \citet{lmsp17} and \citet{dpsf19} have presented RAR observations of ultra-faint dwarf spheroidal galaxies which probe values $\abary\lesssim10^{-12}\msq$ \citep[see][for the corresponding predictions from $\Lambda$CDM simulations]{grpp18}. This, however, is different from  our predictions which hold for the outskirts of much more massive systems and are hence relevant on very different length scales.}

The \emph{bottom panel} of figure~\ref{fig:rar-fullsample} shows the residuals of the RAR ratio data in the \emph{top panel} with the solid purple curve, as a function of baryonic mass $M_{\rm bary}$ (equation~\ref{eq:Mbary-def}). Specifically, on the vertical axis we plot $\log[(\atot/\abary-1)/(\Cal{F}(\abary/a_0)-1)]$ using $\Cal{F}(x)$ from \eqn{eq:RAR-simple} with $a_0=1.2\times10^{-10}\msq$ and $\nu=0.8$. This is conceptually similar to figure~5 of \citet{lmsp17}, who define the residuals using $\log[(\atot/\abary)/(\Cal{F}(\abary/a_0))]$, i.e., without subtracting unity in the numerator and denominator inside the logarithm. This difference is important because the residuals calculated by \citet{lmsp17} will be artificially suppressed in the high-acceleration regime where the numerator and denominator both approach unity. By subtracting this leading behaviour, our definition of the residuals offers a sharper characterisation of the scatter around the median relation. We see from the \emph{bottom panel} of figure~\ref{fig:rar-fullsample} that this scatter is nevertheless small, with a typical value of $\sim0.2$ dex, similar to the scatter seen in $\log[\Delta_a]$ in the \emph{top panel}. We have checked that using the \citet{lmsp17} definition of residuals instead, the typical scatter in the \emph{bottom panel} is even smaller, closer to $\sim0.1$ dex and similar to what they find.

It is clear from the discussion in the Introduction and section~\ref{sec:physics:analytical} that the RAR in our $\Lambda$CDM mocks is an emergent phenomenon rather than a universal law \citep{kw17,desmond17,navarro+17,ludlow+17,tenneti+18}. That discussion also shows that galaxies populating halos of different masses and concentrations might be expected to define different RARs, in general. The RAR is additionally expected to be sensitive to the physics of quasi-adiabatic relaxation of dark matter in the presence of baryons.
In the following subsections, we explore the sensitivity of the RAR to differences in the physical content of galaxies, observational selection  criteria and, importantly, differences in the physical modelling of baryonification. Unless otherwise mentioned, the plots below are formatted identically to the \emph{top panel} of figure~\ref{fig:rar-fullsample}, with the solid and dashed purple curves being repeated from that figure.

\begin{figure}
    \centering
    \includegraphics[width=0.425\textwidth,trim=5 10 5 5,clip]{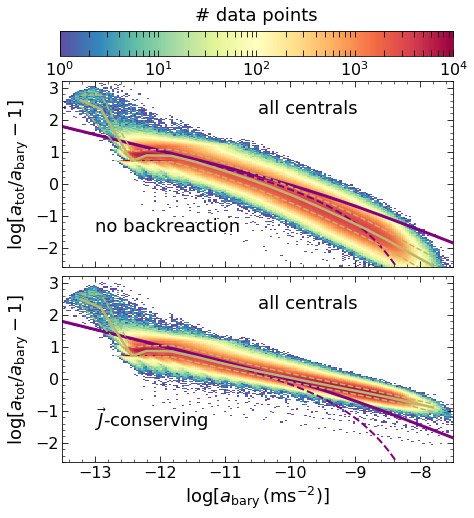}
    \caption{{\bf RAR and relaxation physics.}
    Same as top panel of figure~\ref{fig:rar-fullsample}, but assuming $q_{\rm rdm}=0$ (no baryonic backreaction on the dark matter profile; \emph{top panel}) or $q_{\rm rdm}=1$ (perfect angular momentum conservation; \emph{bottom panel}).  The median and scatter are both different compared to our fiducial choice, $q_{\rm rdm}=0.68$, from figure~\ref{fig:rar-fullsample}, especially in the high-acceleration regime ($\abary\gtrsim10^{-10}\msq$). These trends can be understood analytically (Appendix~\ref{app:analytic}).}
    \label{fig:rar-fullsample-bkrxn}
\end{figure}

\subsection{Sensitivity to relaxation physics}
\label{subsec:relaxation}
\noindent
Figure~\ref{fig:rar-fullsample-bkrxn} shows the effect of changing the details of the quasi-adiabatic relaxation scheme (see the discussion in section~\ref{subsec:xiRAR}). The \emph{top panel} shows the RAR obtained if the baryonic matter had \emph{no backreaction} on the dark matter profile \citep[e.g.,][]{navarro+17}, i.e., setting $q_{\rm rdm}\to0$ in \eqn{eq:X-def} which gives $\xi\to1$ in \eqn{eq:ardmByabary-thiswork} and leads to \eqn{eq:ardmByabary-pert}. The ultra-low-acceleration regime is essentially unchanged as compared to the default case in figure~\ref{fig:rar-fullsample}, which is not surprising since this arises from the outer, dark matter dominated regions of the halo where the dark matter profile is relatively unaffected by the presence of baryons in any case. In the \emph{high-acceleration} regime, on the other hand, we see a dramatic effect: the median RAR is substantially lower, and the scatter is substantially higher, than in the default case.  

The \emph{bottom panel} shows the RAR in the opposite limit where baryonic backreaction perfectly conserves angular momentum, which we model by setting $q_{\rm rdm}=1$ in \eqns{eq:ardmByabary-thiswork} and~\eqref{eq:xi-soln}. As expected, the ultra-low-acceleration regime is unaffected. In the high-acceleration regime, the RAR  is now substantially \emph{higher} than in the default case, with a substantially smaller scatter. Appendix~\ref{app:analytic} provides analytic understanding of the strong dependence on $q_{\rm rdm}$.

Our default choice of $q_{\rm rdm}=0.68$ and, indeed, the choice of functional form in \eqn{eq:X-def} adopted from ST15, is subject to some theoretical uncertainty arising from various choices in modelling baryonic feedback physics (such as winds driven by supernovae or active galactic nuclei) made while performing hydrodynamical simulations. ST15 do not provide any error on the value of $q_{\rm rdm}$ and, more generally, the dependence of quasi-adiabatic relaxation on galaxy and halo properties has also not been systematically studied in the literature to date \citep[although see][for related studies]{cpvh19,cpta21}. Considering this theoretical uncertainty, as well as the sharp sensitivity of the high-acceleration RAR to the physics of quasi-adiabatic relaxation, the good agreement between the default case and the observed relation is truly remarkable, especially since the original ST15 model made \emph{no reference} to the RAR. A different point of view would then be to think of RAR observations in the high-acceleration regime as providing constraints on the value of $q_{\rm rdm}$ (or, more generally, the form of equation~\ref{eq:X-def}). In this context, it is worth noting that the RAR in this regime as defined by spiral galaxies is claimed to be better described by setting $\nu=1$ rather than $\nu=0.8$ in \eqn{eq:RAR-simple} \citep[e.g.][]{efe20}, which would pull the median relation lower and might be better described by decreasing the value of $q_{\rm rdm}$ (see figure~\ref{fig:cubicRAR}). We return to this point below.

\subsection{Sensitivity to baryonic content}
\label{subsec:barycontent}
\noindent
We next investigate the sensitivity of the RAR to the baryonic content of galaxies. We focus here on the presence/absence of an \Hi\ disk, and on the relative contribution of the expelled gas  (`egas') component, which is a proxy for the circum-galactic medium. Galaxies with different `egas' fractions may be expected to behave quite differently in the halo outskirts and consequently in the ultra-low-acceleration regime of the RAR.

\begin{figure}
\centering
\includegraphics[width=0.425\textwidth,trim=5 10 5 5,clip]{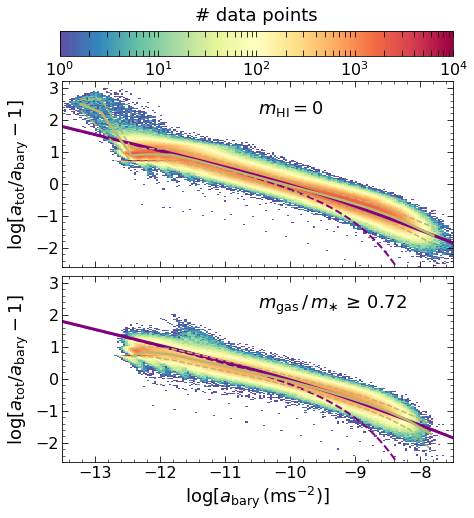}
\caption{{\bf RAR and cold gas content.}
Same as top panel of figure~\ref{fig:rar-fullsample}, but for `bulge-dominated' galaxies having $m_{\Hi}=0$ \emph{(top panel)} or gas-rich `spirals' with $m_{\rm gas}/m_\ast \geq 0.72$ \emph{(bottom panels)}.  Here, $m_{\rm gas}=1.33\,m_{\Hi}$, and the threshold value corresponds to the $95^{\rm th}$ percentile of $m_{\rm gas}/m_\ast$ in the mass range $9.9 \leq \log[m_\ast/(\Mhsq)] \leq 10.9$. 
The break in the median relation at ultra-low accelerations ($\abary\lesssim10^{-12}\msq$) is restricted to bulge-dominated objects.}
\label{fig:rar-coldgas}
\end{figure}

The \emph{top panel} of figure~\ref{fig:rar-coldgas} shows the RAR of mock `bulge-dominated' galaxies without \Hi\ disks, while the \emph{bottom panel} shows the RAR of gas-rich, disk-dominated `spiral' galaxies. We see that bulge-dominated galaxies span a wider range of \abary\ values than the spirals do. In the low-acceleration regime of overlap between the two samples, the median RAR for the two samples is indeed very similar, consistent with the observations quoted above. The  scatter around the median is somewhat smaller for spirals ($\sigma_{\log[\atot]}\sim0.06$ dex) than for bulge-dominated galaxies ($\sim0.075$ dex). 
Interestingly, the break from the MOND-inspired relations at ultra-low accelerations is restricted to the bulge-dominated systems. We discuss this further below.

We have also checked that splitting our default galaxy sample by the value of $M_{\rm bary}$ (with the split defined at the median value $M_{\rm bary}\sim2\times10^{10}\Msun$) leads to results qualitatively very similar to figure~\ref{fig:rar-coldgas}, with the massive sample behaving like the spirals and the low-mass sample behaving like the bulge-dominated galaxies. 
The stark differences between such samples at ultra-low accelerations motivate us to study the effect of the one baryonic component that reaches the halo outskirts, namely the expelled gas (`egas') which we discuss next.

\begin{figure}
\centering
\includegraphics[width=0.425\textwidth,trim=5 10 5 5,clip]{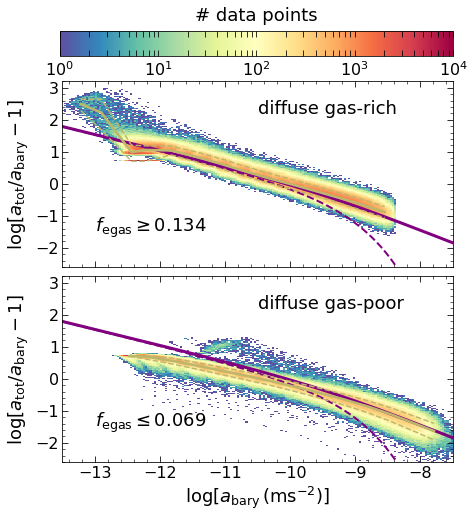}
\caption{{\bf RAR and diffuse gas content.}
    Same as top panel of figure~\ref{fig:rar-fullsample}, but for galaxies with $f_{\rm egas}\geq 0.134$, i.e. rich in expelled or diffuse gas \emph{(top panel)} or $f_{\rm egas}\leq 0.069$, i.e., poor in diffuse gas \emph{(bottom panel)}. 
    The threshold values for rich and poor systems respectively correspond to the $90^{\rm th}$ and $10^{\rm th}$ percentile of $f_{\rm egas}$ in the mass range $9.9 \leq \log[m_\ast/(\Mhsq)] \leq 10.9$. 
    The break in the median relation at ultra-low accelerations ($\abary\lesssim10^{-12}\msq$) is restricted to diffuse gas-rich objects. Section~\ref{subsec:barycontent} argues that $f_{\rm egas}$ is the primary variable controlling the form of the RAR in this regime.}
\label{fig:rar-diffusegas}
\end{figure}

Figure~\ref{fig:rar-diffusegas} shows the RAR for galaxies with large \emph{(top panel)} and small \emph{(bottom panel)} values of the expelled gas mass fraction $f_{\rm egas}$, i.e., galaxies rich and poor, respectively, in diffuse gas content. We see that diffuse gas-rich galaxies tend to populate low and ultra-low accelerations, while diffuse gas-poor galaxies populate high and low accelerations. As compared to the split between bulge-dominated and spiral galaxies in figure~\ref{fig:rar-coldgas}, in this case we see distinct differences between the two samples already at low accelerations $10^{-12}\msq\lesssim\abary\lesssim10^{-11}\msq$, with the median relation of gas-poor galaxies being lower than that of gas-rich galaxies. This indicates that diffuse gas content is more important than morphology in determining the typical RAR at low accelerations. 

We also see that the break from the MOND-inspired relations at ultra-low accelerations is restricted to diffuse gas-rich galaxies. It is not surprising that galaxies with a large amount of  diffuse gas dominate the RAR arising from the halo outskirts, since the `egas' component of our default model is essentially a uniform density sphere at scales $r<R_{\rm vir}$ (see figure~4 of PCS21). In fact, this also explains the results of figure~\ref{fig:rar-coldgas}, since bulge-dominated galaxies with $m_{\Hi}=0$ are likely to have higher values of $f_{\rm egas}$ due to the baryonic mass conservation constraint. 

\begin{figure*}
    \centering
    \includegraphics[width=0.425\textwidth,trim=5 10 5 5,clip]{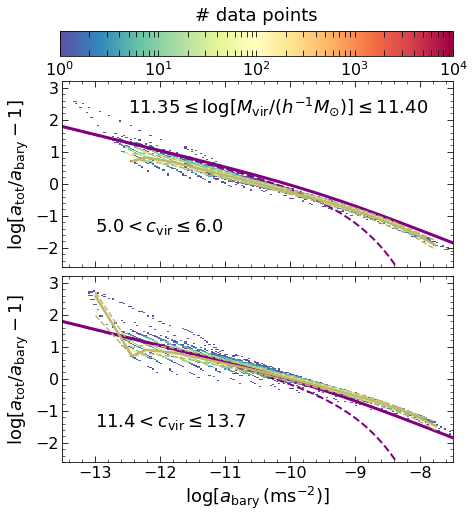}
    \includegraphics[width=0.425\textwidth,trim=5 10 5 5,clip]{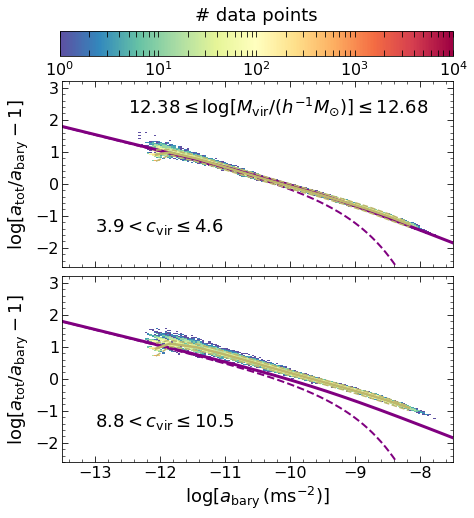}
    \caption{{\bf RAR and halo properties.}
    Same as top panel of figure~\ref{fig:rar-fullsample}, but for centrals with host halo mass $m_{\rm vir}$ and concentration $c_{\rm vir}$ restricted  to narrow ranges. 
    The \emph{left (right) panels} show results for $m_{\rm vir}$ values in the $20$-$25$ ($90$-$95$) percentile range, i.e. for low (high) halo mass. In each bin of $m_{\rm vir}$, the \emph{top (bottom) panel} shows results for the corresponding $10$-$20$ ($80$-$90$) percentile range of $c_{\rm vir}$, i.e. for low (high) concentrations.
    The scatter is very low in each bin of $(m_{\rm vir},c_{\rm vir})$ but the median relation varies systematically with both $m_{\rm vir}$ and $c_{\rm vir}$, consistent with analytical arguments that host halo mass and concentration are the primary variables responsible for the scatter in the RAR of any galaxy sample (see  section~\ref{subsec:xiRAR} and Appendix~\ref{app:analytic}). }
    \label{fig:rar-massconc}
\end{figure*}

The specific form of the sharp break in the RAR from MOND-like predictions at ultra-low accelerations for the high-$f_{\rm egas}$ sample is a consequence of the  choice of spatial distribution of the `egas' component, which is the same as motivated by ST15 in modelling the matter power spectrum of hydrodynamical $\Lambda$CDM simulations. This result, together with the difference between the median RAR of galaxies rich and poor in diffuse gas at accelerations $10^{-12}\msq\lesssim\abary\lesssim10^{-11}\msq$, are testable predictions of the $\Lambda$CDM+baryons framework.

\subsection{Sensitivity to halo mass and concentration}
\label{subsec:massconc}
\noindent
The results in the preceding two subsections focused on the dependence of the median RAR at high and ultra-low accelerations on variations in the  underlying baryon-dark matter response physics and the baryonic content of galaxies. In this subsection, we aim to understand the scatter around the median relation.

As we saw in section~\ref{subsec:xiRAR}, RAR as a function of the relaxation ratio $\xi$  has \emph{zero scatter} in the $\Lambda$CDM+baryons framework. The scatter in the  RAR as a function of \abary\ is therefore entirely due to the scatter between $\xi$ and $\abary$, which in turn is  expected to be driven almost entirely by the variation in halo mass $m_{\rm vir}$ and concentration $c_{\rm vir}$ for galaxies with similar baryonic content. This means, if we focus on galaxies in narrow ranges of $(m_{\rm vir},c_{\rm vir})$, the resulting RAR should have very little scatter but a mean trend that depends on the values of $m_{\rm vir}$ and $c_{\rm vir}$, in general.

We test this idea in figure~\ref{fig:rar-massconc}. 
The \emph{left (right) panels} show the RAR for our default model, with galaxies selected to be in a narrow range of low (high) $m_{\rm vir}$. Within each such range, the \emph{top (bottom) panels} further split the galaxies into narrow ranges of low (high) values of $c_{\rm vir}$. 
It is visually obvious that the RAR in each of these $(m_{\rm vir},c_{\rm vir})$ bins has very low scatter (quantitatively, $\sigma_{\log[\atot]}\sim0.03$-$0.04$), while the  median trends depend significantly on the values of $m_{\rm vir}$ and $c_{\rm vir}$. The median RAR tends to increase in amplitude as $m_{\rm vir}$ increases and, at fixed $m_{\rm vir}$, as $c_{\rm vir}$ increases. That is to say, galaxies in massive, high-concentration halos have a median RAR normalisation that is slightly but significantly higher than that of galaxies in low-mass, low-concentration halos. 
We have checked that the results of using different values of $m_{\rm cvir}$ and $c_{\rm vir}$ lead to smooth extrapolations of these trends (see also figure~\ref{fig:rar-massconc-Rhl}).
Appendix~\ref{app:analytic} provides analytic understanding of these trends.

\subsection{Sensitivity to other details}
\label{subsec:details}
\noindent
The preceding subsections, together with  Appendix~\ref{app:analytic}, give us an essentially complete picture of how the median  RAR and  its scatter emerges from the interplay between halo properties, their scalings with baryonic content and the direct cross-talk between baryons and dark matter through quasi-adiabatic relaxation. In this subsection, we explore a few more aspects of the RAR, including its sensitivity to the shape of the `un-baryonified'  dark matter profile, some of the scaling relations underlying our baryonification scheme and technical choices in sampling the rotation curve data. We also show how the RAR responds to systematic changes in optical sample selection for an SDSS-like galaxy sample.

\subsubsection{Dark matter profile}
\label{subsubsec:dmprof}
\noindent
Our default model uses the NFW form to model the initial, `un-baryonified' dark matter profile. We have checked that using an appropriately matched Einasto profile instead \citep{einasto65,cpt05,retana-montenegro12,dm14,klypin+16} leads to essentially no change in the median RAR or its scatter. 
In other words, while the RAR is sensitive to the overall mass and concentration of halos (section~\ref{subsec:massconc}), it is relatively insensitive to changes in the inner and outer slope of the initial dark matter profile.

\begin{figure}
\centering
\includegraphics[width=0.425\textwidth,trim=5 10 5 5,clip]{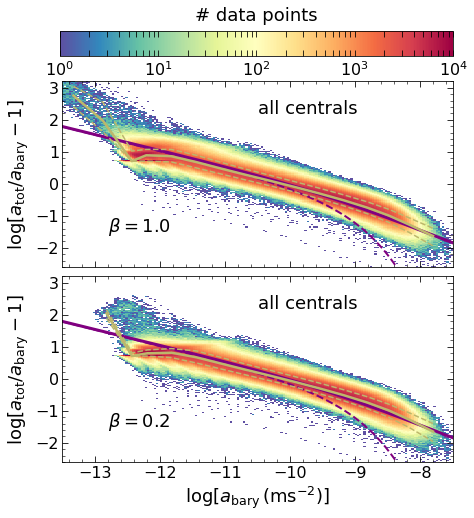}
\caption{{\bf RAR and bound gas fraction.}
Same as top panel of figure~\ref{fig:rar-fullsample}, but with the
baryonification parameter $\beta$ defining the bound gas fraction in \eqn{eq:fbgas} increased \emph{(top panel)} and decreased \emph{(bottom panel)} by $\pm 0.4$ from the default value of $0.6$. The ultra-low-acceleration regime ($\abary\lesssim10^{-12}\msq$) responds to low values of $f_{\rm bgas}$. The text relates this to the relative mass fractions and hence spatial distributions of bound and diffuse gas in the halo outskirts (section~\ref{subsubsec:galscaling:boundgas}).}
    \label{fig:rar-fullsample-beta}
\end{figure}

\begin{figure*}
    \centering
    \includegraphics[width=0.425\textwidth,trim=5 10 5 5,clip]{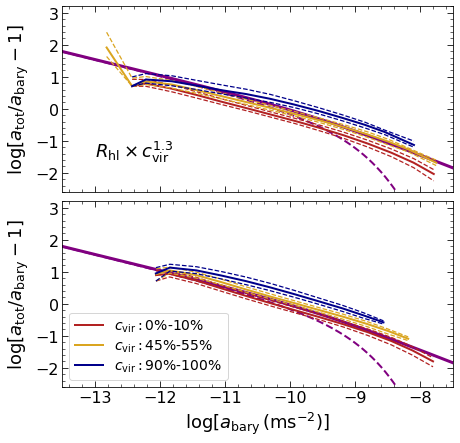}
    \includegraphics[width=0.425\textwidth,trim=5 10 5 5,clip]{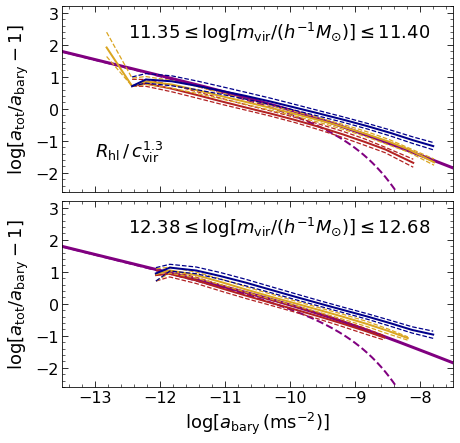}
    \caption{{\bf RAR and stellar profile.}
    Similar to figure~\ref{fig:rar-massconc}, showing results when the baryonification parameter $R_{\rm hl}/R_{\rm vir}$ (whose default value is $0.015$) is multiplied \emph{(left panels)} or divided \emph{(right panels)} by $(c_{\rm vir}/\avg{c_{\rm vir}|m_{\rm vir}})^{1.3}$, with $\avg{c_{\rm vir}|m_{\rm vir}}$ being the median concentration at fixed halo mass, which leads to a scatter of $\sim0.2$dex in $R_{\rm hl}/R_{\rm vir}$ around a median value of $0.015$ in each case. For this figure, we show only the median and central $68\%$ scatter of the RAR for centrals selected by $m_{\rm vir}$ and $c_{\rm vir}$: \emph{top (bottom) panels} correspond to halos in the 20-25 (90-95) percentile ranges of $m_{\rm vir}$, and the coloured lines in each panel further split the samples into the indicated percentiles of $c_{\rm vir}$ (c.f. figure~\ref{fig:rar-massconc}). While galaxies at fixed $(m_{\rm vir},c_{\rm vir})$ trace out the same RAR in each case, the range of \abary\ explored depends sensitively on whether the correlation between $R_{\rm hl}$ and $c_{\rm vir}$ is positive or negative. See  section~\ref{subsubsec:galscaling:stellar} for a discussion.}
    \label{fig:rar-massconc-Rhl}
\end{figure*}

\subsubsection{Baryonic scaling relations: bound gas fraction}
\label{subsubsec:galscaling:boundgas}
\noindent
The mass fraction $f_{\rm bgas}$ in hot, bound gas in our default model from PCS21 is the same as used by ST15 and is given by
\be
f_{\rm bgas} = (\Omega_{\rm b}/\Omega_{\rm m})\times\left[1+(M_{\rm c}/m_{\rm vir})^\beta\right]^{-1}\,,
\label{eq:fbgas}
\ee
with $M_{\rm c}=1.2\times10^{14}\Mh$ and $\beta=0.6$. ST15 showed that there is considerable room for variation in the values of $M_{\rm c}$ and especially $\beta$ when considering the effects of baryonification on the matter power spectrum alone. Moreover, as discussed by PCS21, the relation above has been extrapolated to halos  with $m_{\rm vir}\sim10^{11}\Mh$ in our  mocks, well below the scale $m_{\rm vir}\gtrsim10^{13}\Mh$ at which ST15 calibrated their results. It is therefore interesting to ask how  the RAR is affected by variations in these model parameters. We explore this in figure~\ref{fig:rar-fullsample-beta}, focusing  on $\beta$ since the pivot scale $M_{\rm c}$ is reasonably well constrained by the X-ray cluster observations cited by ST15. We see that variations in $\beta$ primarily affect the ultra-low-acceleration regime, changing the slope of the median RAR. This is sensible, because an increase in $f_{\rm bgas}$ at the mass scales of our interest (by decreasing $\beta$) will correspondingly decrease $f_{\rm egas}$ due to baryonic mass conservation and hence change the relative spatial behaviour of the `bgas' and `egas' components  in the halo outskirts (see figure~\ref{fig:relaxation-ratio}). Thus, the ultra-low-acceleration regime of the RAR is, in principle, sensitive to the physics of both hot and cold gas in the outer halo.

\subsubsection{Baryonic scaling relations: stellar size}
\label{subsubsec:galscaling:stellar}
\noindent
Our default model treats the stellar profile as a bulge with half-light radius $R_{\rm hl} = 0.015 R_{\rm vir}$, which is approximately the result obtained by \citet{kravtsov13} using a power-law fit to $\sim180$ galaxies. We have not included the scatter of $\sim0.2$ dex around this relation which was reported by \citet{kravtsov13} and which, as discussed by him, could in principle be linked to the halo spin using the formalism of \citet{mmw98}. 
We assess the potential effect of this scatter in figure~\ref{fig:rar-massconc-Rhl}, using halo concentration as a proxy for internal halo properties (the analytical model in Appendix~\ref{app:analytic} suggests that the RAR ought to be sensitive  to a correlation between $R_{\rm hl}/R_{\rm vir}$ and $c_{\rm vir}$).

Halo concentrations in our mocks have a Lognormal distribution at  fixed mass, with a median $\avg{c_{\rm vir}|m_{\rm vir}}$ and a scatter $0.16$ dex taken from \citet{dk15}.
The \emph{left (right) panels} of figure~\ref{fig:rar-massconc-Rhl} show the RAR after multiplying (dividing) the default $R_{\rm hl}/R_{\rm vir}$ for each galaxy by $(c_{\rm vir}/\avg{c_{\rm vir}|m_{\rm vir}})^{\mu}$, thus leading to a positive (negative) correlation between stellar bulge size and initial halo concentration. In this toy model, the entire variation in $R_{\rm hl}/R_{\rm vir}$ is explained by halo concentration; more realistic models would allow room for other variables (such as halo angular momentum, or some unspecified source of stochasticity) to also play a role.
By construction, the modified set of $R_{\rm hl}/R_{\rm vir}$ values obey a Lognormal distribution at fixed halo mass, with median $0.015$ and a scatter of $0.16\times\mu$ dex. We therefore set $\mu=1.3$, which gives a scatter of $\simeq0.2$ dex in $R_{\rm hl}/R_{\rm vir}$ for both choices of the correlation, consistent with \citet{kravtsov13}. 

The \emph{top (bottom) panels} of figure~\ref{fig:rar-massconc-Rhl} use the same ranges of $m_{\rm vir}$ shown in the left and right panels, respectively, of  figure~\ref{fig:rar-massconc}. Similarly to that figure, we further split these fixed-$m_{\rm vir}$ samples into narrow ranges of $c_{\rm vir}$. For this figure alone, so as to highlight differences between the subsamples, we only show the median and central $68\%$ of each RAR using the differently coloured lines. As expected from the discussion in  sections~\ref{subsec:xiRAR} and~\ref{subsec:massconc}, galaxies at fixed $m_{\rm vir}$ and $c_{\rm vir}$  trace the same RAR \emph{regardless} of the sign of the bulge size-halo concentration correlation. We do see a very interesting trace of this signature however, in that the range of values of \abary\ explored by any sample responds sensitively to whether the correlation is positive or negative. In the former case, galaxies in high-concentration halos explore lower values of \abary\ than low-concentration ones, and vice-versa for a negative correlation. This trend can be understood as follows. Consider a specific galaxy with stellar mass $m_\ast$ in an $(m_{\rm vir},c_{\rm vir})$ host. For a positive correlation, a large $c_{\rm vir}$ implies a larger $R_{\rm hl}$ for this galaxy than in the absence of the  correlation. Since the (now flatter) stellar density profile must enclose the same $m_\ast$ inside the same $R_{\rm vir}$, its inner parts are forced to be lower, thus contributing less to $m_{\rm cgal}(<r)$ and hence \abary\ in the inner region, than in the absence of the positive correlation. A negative correlation between $R_{\rm hl}$ and $c_{\rm vir}$ has exactly the opposite effect.

While this shows that there is clearly no new physics explored by such a correlation beyond the dependence of the RAR on $m_{\rm vir}$ and $c_{\rm vir}$ through the relaxation ratio $\xi$, it does lead to a curious degeneracy. It is clear from the left hand panels of figure~\ref{fig:rar-massconc-Rhl} that the RAR obtained from averaging over all $c_{\rm vir}$ values will tend to curve downwards at large \abary\ in the case of a positive $R_{\rm hl}$-$c_{\rm vir}$ correlation.  Further integration over $m_{\rm vir}$ will not change this curvature, so the resulting RAR will be qualitatively similar to that in which there is no $R_{\rm hl}$-$c_{\rm vir}$ correlation but $q_{\rm rdm}$ is smaller (compare top panel of figure~\ref{fig:rar-fullsample-bkrxn}).  Conversely, the right hand panels of figure~\ref{fig:rar-massconc-Rhl} show that an $R_{\rm hl}$-$c_{\rm vir}$ anti-correlation will result in an RAR that would imply a larger $q_{\rm rdm}$ if one assumed there was no $R_{\rm hl}$-$c_{\rm vir}$ correlation (e.g. bottom panel of figure~\ref{fig:rar-fullsample-bkrxn}).  
That curvature in the RAR may arise from the $R_{\rm hl}$-$c_{\rm vir}$ relation rather than $q_{\rm rdm}$ must be kept in mind during any analysis which aims to probe the physics of quasi-adiabatic relaxation using the RAR. 

We noted earlier that setting $\nu=1$ rather than $\nu=0.8$ in \eqn{eq:RAR-simple} might provide a better description of the RAR of spiral galaxies than that of ellipticals. The above discussion shows that our model is capable of explaining such a difference either by decreasing the value of the relaxation parameter $q_{\rm rdm}$, or making $R_{\rm hl}$ correlate positively with $c_{\rm vir}$, or a combination of the two. This is subject to the caveat that the relaxation model in \eqn{eq:X-def} (adopted from ST15) is itself approximate, and we also have not yet self-consistently modelled stellar disks. Using a more physically motivated `size-mass' correlation, such as the one between galaxy size and halo angular momentum alluded to above, can potentially add another dimension to such degeneracies \citep[see also the discussion in][]{desmond17}.  It will be interesting to study such effects in hydrodynamical simulations of cosmological volumes, which we leave to future work.\footnote{We have checked that the same exercise performed using the \Hi\ disk size $h_{\Hi}$ rather than the stellar bulge size $R_{\rm hl}$ leads to no significant effect on the RAR. This is likely because observations allow a scatter of only $\simeq0.06$ dex in $h_{\Hi}$ at fixed $m_{\Hi}$, which leaves room for only a weak correlation, at best, between $h_{\Hi}$ and $c_{\rm vir}$.}

\begin{figure}
\centering
\includegraphics[width=0.425\textwidth,trim=5 10 5 5,clip]{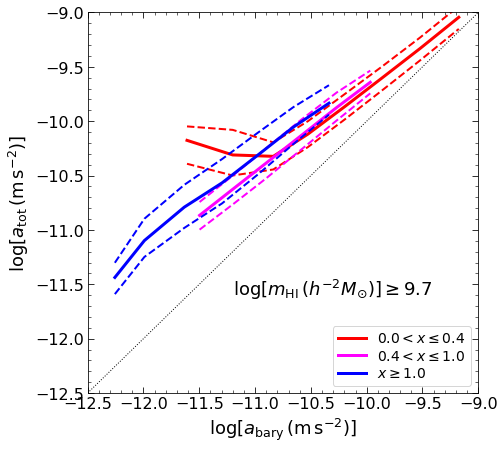}
\caption{{\bf RAR and optical profile.} Dependence of the RAR on $x\equiv r/R_{\rm opt}$, for massive \Hi\ galaxies (for which our sample is volume complete), shown in the format used by \citet{dpsf19} in their analysis of dwarf disk spirals. There is a hint of a dependence on $x$ at low accelerations, in qualitative agreement with the observations. The  upturn at low \abary\ in the low $x$ systems deserves further study.}
\label{fig:dpsf}
\end{figure}

Finally, our discussion of scalings with optical size has an interesting connection to other recent work.  In their study of the RAR in dwarf disk spirals and low surface brightness galaxies, \cite{dpsf19} found that the RAR depends on a third parameter, $x\equiv r/R_{\rm opt}$, where $R_{\rm opt}$ is the scale which contains 83\% of the stellar light.  While this trend is most obvious at accelerations which are smaller than where our mocks are complete, it is plausible that our mocks exhibit something similar, since $R_{\rm opt}$ scales with $R_{\rm hl}$, which scales with halo mass, and we do expect weak trends with halo mass.  Figure~\ref{fig:dpsf} indeed shows a qualitatively similar trend at least at $x\gtrsim0.4$, even though we have made no effort to identify dwarf disk spirals in the mocks.  Performing a more careful comparison would be interesting -- especially for what it may teach us about the interplay between size-mass-angular momentum correlations and $q_{\rm rdm}$ at low accelerations -- but is beyond the scope of this work.  

\begin{figure*}
    \centering
    \includegraphics[width=0.425\textwidth,trim=5 10 5 5,clip]{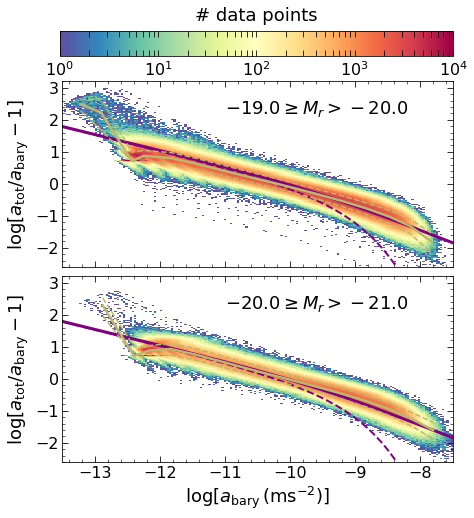}
    \includegraphics[width=0.425\textwidth,trim=5 10 5 5,clip]{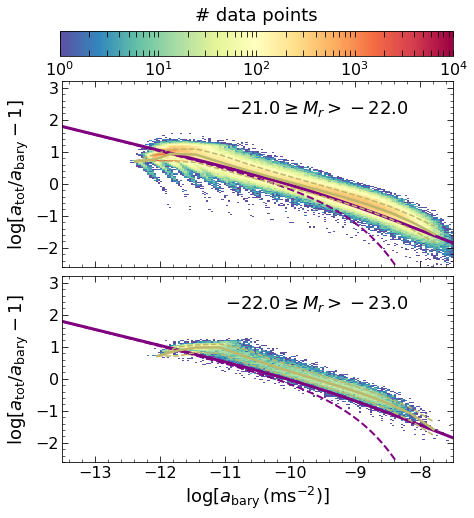}
    \caption{{\bf RAR and luminosity.}
    Same as top panel of figure~\ref{fig:rar-fullsample}, for centrals selected by $r$-band absolute magnitude as indicated, with the samples increasing in luminosity from \emph{top left} $\to$ \emph{bottom left} $\to$ \emph{top right} $\to$ \emph{bottom right}. There is a clear luminosity dependence, with the RAR shifting vertically upwards for brighter samples. This is a natural consequence of the halo mass dependence seen in figure~\ref{fig:rar-massconc} (see section~\ref{subsubsec:lumcolour}).}
    \label{fig:rar-luminosity}
\end{figure*}

\begin{figure*}
    \centering
    \includegraphics[width=0.425\textwidth,trim=5 10 5 5,clip]{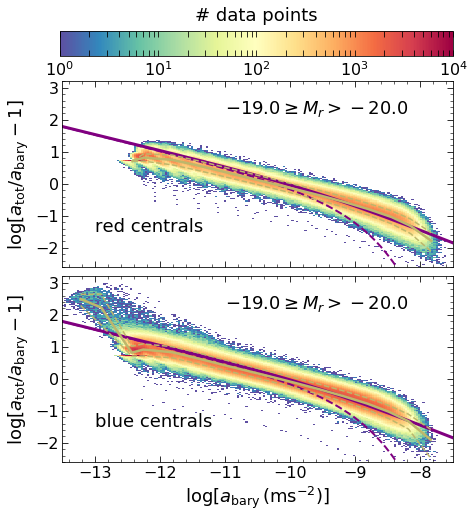}
    \includegraphics[width=0.425\textwidth,trim=5 10 5 5,clip]{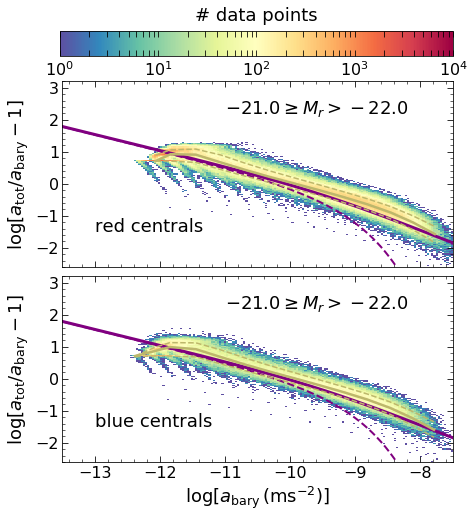}
    \caption{{\bf RAR and optical colour.}
    Same as top panels of figure~\ref{fig:rar-luminosity}, with the \emph{top (bottom) panels} now focusing on red (blue) galaxies. The separation between red and blue uses the luminosity-dependent threshold on $g-r$ colour given by \eqn{eq:gr-cut}. There is a sharp difference between red and blue centrals in the fainter bin at ultra-low accelerations ($\abary\lesssim10^{-12}\msq$, c.f. figure~\ref{fig:rar-diffusegas}), with relatively little difference between the two samples in the brighter bin. These trends can be traced back to differences, or lack thereof, in the mass fraction  $f_{\rm egas}$ of diffuse gas in these samples (see section~\ref{subsubsec:lumcolour}).}
    \label{fig:rar-colour}
\end{figure*}

\subsubsection{Optical selection}
\label{subsubsec:lumcolour}
\noindent
Observational analyses of the RAR are typically limited by the quality of rotation curve \citep[e.g.,][]{lms16b} or velocity dispersion \citep[e.g.,][]{cbsg19} measurements, which can introduce inhomogeneities  in the statistical properties of the associated galaxy sample. Since our mock catalogs have `perfect' rotation curve measurements, we can use them to ask how the RAR responds to systematic variations in, say, optical sample selection for SDSS-like galaxies.

Figure~\ref{fig:rar-luminosity} shows the RAR using our default model for mock centrals chosen to lie in bins of luminosity (one bin in each panel), represented by $r$-band absolute magnitudes $M_r$ (see PCS21 for a detailed definition). The samples increase in luminosity going from \emph{top left} $\to$ \emph{bottom left} $\to$ \emph{top right} $\to$ \emph{bottom right}, with the faintest bin corresponding to sub-$L_\ast$ centrals (median $m_{\rm vir}\simeq10^{11.6}\Mh$) and the brightest to BCGs of massive clusters (median $m_{\rm vir}\simeq10^{13.8}\Mh$). We see a clear indication that the normalisation of the median RAR increases with increasing luminosity, which is sensible given the results of figure~\ref{fig:rar-massconc} and the fact that central luminosity correlates positively with halo mass in our mocks. Interestingly, recent results suggest that the RAR of observed galaxy clusters also has an elevated normalisation relative to that of galaxy samples (\citealp{tian+20}; see also \citealp{pgsd21}), in qualitative agreement with our results.
We also find that the typical scatter of the RAR
varies non-monotonically with luminosity, being $\sigma_{\log[\atot]}\sim 0.067, 0.069,0.077,0.066$ in successively brighter bins.

It is also interesting to split the galaxy sample at fixed luminosity by colour. We use a luminosity-dependent threshold on the $g-r$ colour index of each of our mock central galaxies, given by \citep{zehavi+11}
\be
(g-r)_{\rm cut}(M_r) \equiv 0.21 - 0.03M_r\,.
\label{eq:gr-cut}
\ee
We classify galaxies having  $g-r \geq (g-r)_{\rm cut}(M_r)$ as `red' and the rest as `blue'.
Figure~\ref{fig:rar-colour} shows the resulting RAR for two of the luminosity bins shown in figure~\ref{fig:rar-luminosity}. In the absence of `beyond halo mass' effects such as galactic conformity, galaxy colours in our default mocks correlate only with galaxy luminosity, not with halo mass or concentration. Naively, therefore, we should not expect \emph{any} difference in the RAR of red and blue galaxies at fixed luminosity. This  is  indeed the  case for the brighter luminosity bin shown in the \emph{right panels} of figure~\ref{fig:rar-colour}. There is, however, a secondary correlation one must account for in a luminosity-complete sample. This is the fact  that, due to a colour-dependent mass-to-light ratio, blue centrals of a given luminosity will have lower stellar masses $m_\ast$ than red centrals with similar luminosity \citep[see, e.g., figure~3 of PCS21 and figure~4 of][]{pkhp15}. The decrease in $m_\ast$ from red to blue objects is accompanied by an increase in $f_{\rm egas}$ due to baryonic mass conservation. We saw already, in figure~\ref{fig:rar-diffusegas}, that samples with higher $f_{\rm egas}$ tend to break away from the smooth, MOND-inspired RAR functional forms at  ultra-low accelerations. Figure~\ref{fig:rar-massconc} also showed that this break is prominent only for galaxies with low-mass hosts, whose median RAR can reach the ultra-low-acceleration regime.
Not surprisingly, then, we see in the \emph{left panels} of figure~\ref{fig:rar-colour} that faint blue centrals trace out exactly the same break, which is correspondingly absent for faint red objects. At higher luminosity, the corresponding halo masses are higher, so that the median RAR does not reach the ultra-low acceleration regime, leading to identical RARs for red and blue galaxies as discussed above.

\begin{figure}
\centering
\includegraphics[width=0.425\textwidth,trim=5 10 5 5,clip]{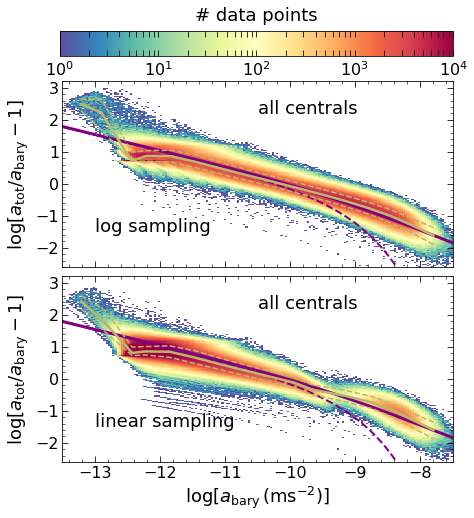}
\caption{{\bf RAR and rotation curve sampling.}
\emph{(Top panel:)} Identical to top panel of figure~\ref{fig:rar-fullsample}, i.e. using rotation curves sampled with 20 logarithmically spaced points between $(0.001,1)\times R_{\rm vir}$ for each galaxy. \emph{(Bottom panel:)} Same as top panel, but sampling each rotation curve using 40 linearly spaced points between $(0.001,1)\times R_{\rm vir}$. 
The density of points in different parts of the RAR depends strongly on the sampling of the rotation curves, but the median relation and its scatter are relatively insensitive to this choice.}
\label{fig:rar-sampling}
\end{figure}

\begin{figure*}
    \centering
    \includegraphics[width=0.425\textwidth,trim=5 8 5 5,clip]{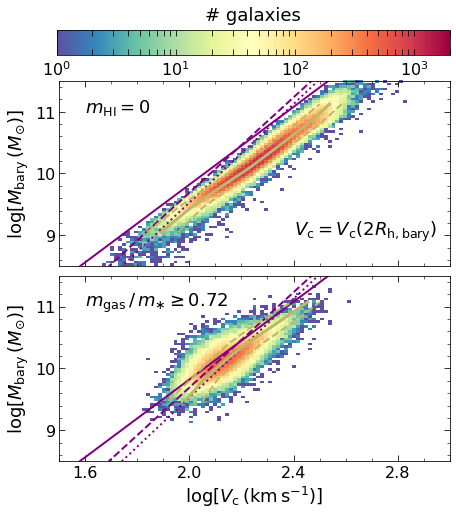}
    \includegraphics[width=0.425\textwidth,trim=5 8 5 5,clip]{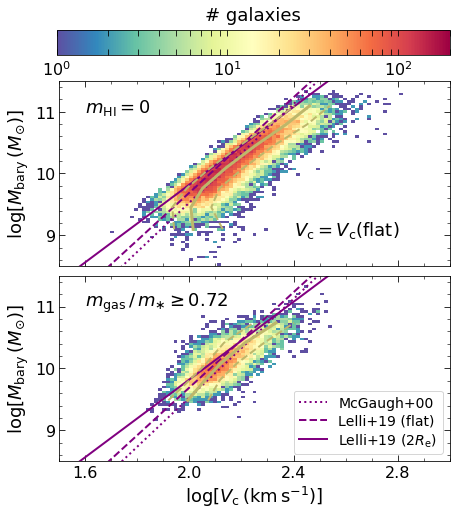}
    \caption{{\bf Baryonic Tully-Fisher relation (BTFR)} using the default baryonification model. Coloured histograms show the distribution of $M_{\rm bary}$, the `baryonic' (i.e., stellar + cold gas) mass from \eqn{eq:Mbary-def} against circular velocity $V_{\rm c}$ defined in different ways. \emph{Left panels} use $V_{\rm c}$ measured at $r=2R_{\rm h,bary}$ (where $R_{\rm h,bary}$ is the baryonic half-mass radius), while \emph{right panels} use $V_{\rm c}$ calculated as the mean  circular velocity in the flat part of the rotation curve using the algorithm of \citet{lms16a}. 
    Note the difference in the range of the colour bars in the left and right panels.
    \emph{Top panels} show `bulge-dominated' galaxies with $m_{\Hi}=0$, while \emph{bottom panels} show gas-rich `spiral' galaxies with $m_{\rm gas}/m_\ast\geq0.72$ (here, $m_{\rm gas}=1.33\,m_{\Hi}$). The threshold value is chosen as described in the  caption of figure~\ref{fig:rar-coldgas}. 
    Solid yellow curve in each panel shows the median $V_{\rm c}$ in bins of baryonic mass. Dashed yellow curves show the corresponding $16^{\rm th}$ and $84^{\rm th}$ percentiles (i.e., the horizontal scatter). For comparison, the purple lines (repeated in each panel) show the observed relations using $V_{\rm c}$ measured in the flat part of the rotation curve from \citet[][dotted: $M_{\rm bary}\propto V_{\rm c}^{3.98}$]{msbdb00} and \citet[][dashed: $M_{\rm bary}\propto V_{\rm c}^{3.85}$]{lelli+19} and using $V_{\rm c}$ measured at twice the observed half-light radius from \citet[][solid: $M_{\rm bary}\propto V_{\rm c}^{3.14}$]{lelli+19}.
    We see that spirals in the mock catalog have steeper BTFR slopes as well as higher normalisations than bulge-dominated galaxies, with the effect being more pronounced when $V_{\rm c}$ is measured in the flat part of the rotation curve.
    }
    \label{fig:btfr}
\end{figure*}

\begin{figure}
    \centering
    \includegraphics[width=0.475\textwidth]{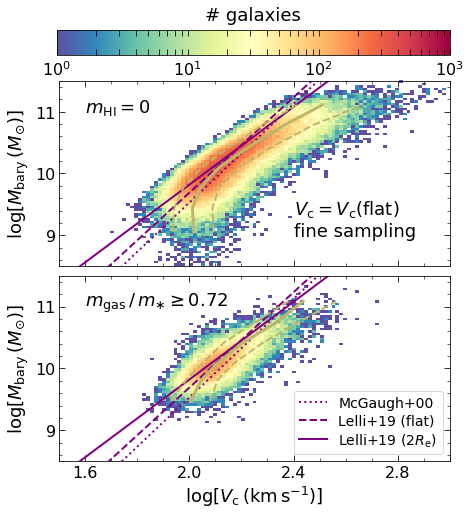}
    \caption{{\bf BTFR and rotation curve sampling.}
    Same as right panels of figure~\ref{fig:btfr}, except that the rotation curves were sampled with twice the number of points. We see that the BTFR, especially of massive spirals, is shallower than in figure~\ref{fig:btfr}, and many more galaxies are included in the relation. The BTFR in the flat part of the rotation curve is hence sensitive to sampling choices.
    }
    \label{fig:btfr-sampling}
\end{figure}

\subsubsection{Sampling}
\label{subsubsec:sampling}
\noindent
All our results above have been based on an arbitrarily chosen sampling of the rotation curve of each galaxy, using $20$ logarithmically spaced values of $r$ in the range $(0.001,1)\times R_{\rm vir}$ (section~\ref{subsec:default}). Since observed rotation curves are typically inhomogeneous in the available sampling \citep[e.g.,][]{lmsp17}, it is important to check what role sampling plays in establishing the median RAR and its scatter. We test this in figure~\ref{fig:rar-sampling} by comparing our default results with those obtained using a different sampling choice, now using $40$ \emph{linearly} spaced points in the same range $(0.001,1)\times R_{\rm vir}$ for each galaxy. Visually, the resulting histogram is very different from the default case, being over-sampled at ultra-low accelerations and under-sampled at high accelerations (as expected from the fact that the linear sampling decreases the number of available points in the inner halo). Encouragingly, though, the median RAR as well as the scatter are relatively unaffected across the entire range of \abary\ (nearly 6 orders of magnitude) probed in the plot. We conclude that sampling choices are not expected to be a major source of systematic uncertainty in the median and scatter of the RAR.

\section{Baryonic Tully-Fisher relation}
\label{sec:btfr}
\noindent
The RAR is closely linked with the so-called baryonic Tully-Fisher relation \citep[BTFR,][]{msbdb00} $M_{\rm bary} \propto V_{\rm c}^\alpha$, which generalises the classical Tully-Fisher relation $m_\ast\propto V_{\rm c}^\alpha$ \citep{tf77} to include the mass in cold gas in addition to stellar mass. The BTFR with a slope $\alpha=4$ and a small scatter $\sim0.1\,{\rm dex}$ has been shown to be valid over a wide dynamic range of baryonic mass for gas-rich spiral galaxies \citep[e.g.,][although see below]{lms16a}.

The quantity $V_{\rm c}$ in the BTFR is an estimate of the circular velocity in the outer parts of the galaxy and is meant to be a proxy for the total matter content of each system. 
The precise definition of $V_{\rm c}$ has been the subject of some discussion, and the inferred slope $\alpha$ of the BTFR is rather sensitive to the assumed definition of $V_{\rm c}$ \citep{bgvdb16}. Defining $V_{\rm c}$ at some fixed multiple of the disk scale length typically leads to $\alpha\simeq3$, while defining $V_{\rm c}$ in the `flat part' of the rotation curve (see below), typically yields steeper slopes  $\alpha\simeq4$ \citep{bss16,lelli+19}. It has been argued \citep{lelli+19} that the latter must be more fundamental, since the associated scatter in the BTFR is reduced as compared to that when using $V_{\rm c}$ tied to the disk scale length. It has been further argued \citep{whd19}, that the RAR in the  low-acceleration regime $10^{-12}\msq\lesssim\abary\lesssim10^{-10}\msq$ is a simple algebraic consequence of a BTFR with slope $\alpha=4$, so that models which satisfy the BTFR with this slope are guaranteed to follow the observed RAR at low accelerations.  

In this section, we use our mock galaxies to place the above results in the context of the analytical and numerical arguments concerning the RAR from the preceding sections. Figure~\ref{fig:btfr} shows the BTFR for our mock centrals using the default baryonification scheme and the same definition of $M_{\rm bary}$ (equation~\ref{eq:Mbary-def}) used in figure~\ref{fig:rar-coldgas}. 

The \emph{top panels} of the figure focus on pure bulge-like galaxies while the \emph{bottom panels} show results for gas-rich, disk-dominated systems, with the split being identical to the one used in figure~\ref{fig:rar-coldgas}. The \emph{left panels} show results when  $V_{\rm c}$ is defined as $V_{\rm c} = v_{\rm rot}(r=2R_{\rm h,bary})$, where $R_{\rm h,bary}$ is the baryonic half-mass radius, i.e. at the  same location as used  to calculate $M_{\rm bary}$. In the \emph{right panels}, we follow \citet{lms16a} and discard galaxies for which the `outermost' part of the  rotation curve is either  rising or falling too steeply. We pick $r=0.3R_{\rm vir}\simeq20R_{\rm hl}$ as the outermost measured radius  and define the threshold steepness by requiring that successive residuals between $v_{\rm rot}(r)$ at smaller radii and  the mean $v_{\rm rot}$ in the outermost region be  smaller than $2\%$. In other words, we implement the iterative algorithm of \citet{lms16a} with a threshold of $2\%$ instead of the $5\%$ those authors used. We only use galaxies with at least 3 usable values of $r$, which was also done by \citet{lms16a}. Another detail is that we perform this exercise on a linearly sampled grid of $r$ values containing 6 points between $(0.01,0.3)\times R_{\rm vir}$. The resulting mean value of $v_{\rm rot}$ is then an estimate of $V_{\rm c}(\rm flat)$, the circular velocity in the `flat part' of the rotation curve. The sample of galaxies selected by this analysis is an order of magnitude smaller than the one used in the \emph{left panels}.

We will shortly discuss the dependence of our results on the (admittedly arbitrary) technical choices in measuring $V_{\rm c}(\rm flat)$. We first note, however, that each of our mock samples defines a reasonably tight BTFR in figure~\ref{fig:btfr}.\footnote{Due to their luminosity-complete nature, our mocks are complete in stellar and \Hi\ mass only for thresholds $m_\ast\gtrsim10^{9.85}\Mhsq$ and $m_{\Hi}\gtrsim10^{9.7}\Mhsq$ (see PCS21 for details), which leads to a somewhat complicated completeness threshold as a function of $M_{\rm bary}$. In order to avoid the resulting Malmquist bias effects in characterising the BTFR, throughout this section we report results in bins of $M_{\rm bary}$ rather than $V_{\rm c}$. See \citet{bgvdb16,lelli+19} for a discussion of the complications in fitting BTFR slopes to observed data which, in addition to selection effects, also have errors on both variables.} 
A closer comparison with results from the literature (purple lines) shows that (a) the BTFR of pure bulges has a slope close to $\alpha\simeq3$, decidedly shallower than that of gas-rich spirals which are closer to $\alpha\simeq4$. Focusing on the latter (i.e., the lower panels), we also see some hint at the highest masses that using $V_{\rm c}({\rm flat})$ leads to a slightly steeper slope than $V_{\rm c}(2R_{\rm h,bary})$. The horizontal scatter around the median relation is $\sim0.055\,{\rm dex}$.

These trends are easily understood. For all objects, both the stellar mass and $V_{\rm c}(\rm flat)$ are tightly correlated with halo mass.  For bulges, $m_{\rm bary}$ on the relevant scales is dominated by the stellar component, so the curvature in the top panel, which results in a shallower effective slope, is a consequence of the curvature in the $m_*$-$m_{\rm halo}$ relation.  Spirals in our mocks have the {\em same} $m_*$-$m_{\rm halo}$ relation, but because our mocks have \Hi\ gas fractions decreasing with mass, spirals do not probe the higher halo masses where the curvature matters.  This is why their $m_*-V_{\rm c}$ relation appears to be steeper.  Adding the \Hi\ mass to $m_*$, so as to obtain $m_{\rm bary}$, lifts the relation for spirals above that for the bulges, bringing them closer to the observed BTFR. 

Thus far, \emph{our default BTFR results are in reasonable agreement with observations.} This is already interesting, because a comparison with figure~\ref{fig:rar-coldgas} shows that, although gas-rich spirals with $\alpha\simeq4$ do fall on the observed RAR in the low-acceleration regime, \emph{so do pure bulges with $\alpha\simeq3$}. In other words, while being on the BTFR may guarantee being  on the RAR \citep{whd19}, the RAR is obeyed by a \emph{much wider} class of galaxies. This implies, firstly, that statements such as `the RAR is a natural consequence of the BTFR', which suggest that the BTFR is more fundamental than the RAR, must  be treated with caution. Secondly, the converse is \emph{also} not true in our mocks: galaxies which contribute to the low-acceleration RAR \emph{need not} obey the BTFR with $\alpha=4$ (compare the upper panels of figures~\ref{fig:rar-coldgas} and~\ref{fig:btfr}), in contrast with some claims in the literature \citep[see, e.g., the discussion in section 7.1 of][]{lmsp17}.

Things become even more interesting when one starts to question the various technical choices used in defining `good' rotation curves. Figure~\ref{fig:btfr-sampling} shows the results of an exercise identical to the one described above in estimating $V_{\rm c}({\rm flat})$, with the only difference being that we now used a linearly spaced array of $r$ values with $12$ instead of $6$ points in the range $(0.01,0.3)\times R_{\rm vir}$, \emph{without changing} the flatness threshold of $2\%$. It is obvious, upon some thought, that this change will relax the flatness restriction and allow more galaxies to be used in the sample containing valid $V_{\rm c}({\rm flat})$ values. We see that the resulting BTFR for both bulges as well as spirals are now very different from those in the right panels of figure~\ref{fig:btfr}. In particular, the BTFR of spirals is \emph{now consistent with $\alpha\simeq3$}. 
We have checked that similar results are obtained upon relaxing the flatness threshold to $5\%$ for our default sampling, as well as when modifying the definition of the `flat part' to use analytical slopes $\der\ln v_{\rm rot}/\der\ln r$ in constraining the degree of flatness.
Considering the lack of homogeneity of rotation curve sampling in (otherwise very high quality) data-bases such as SPARC \citep{lms16b} which has been used in many recent BTFR analyses, our mock results call for a great deal of caution in interpreting a BTFR analysis in the context of  competing gravitational theories.

In contrast, the differences seen in the median RAR and its scatter in figure~\ref{fig:rar-sampling} due to (rather dramatic) changes in sampling the rotation curves are relatively minor in comparison. The RAR is therefore a much more observationally robust probe of the nature of gravity at galactic scales than is the BTFR.

\section{Conclusions}
\label{sec:conclude}
\noindent
We have presented new analytical insights into the structure and origin of the radial acceleration relation (RAR) between the total (\atot) and baryonic (\abary) centripetal acceleration profiles of galaxies in the $\Lambda$CDM framework. 

Our key result follows from the realisation (section~\ref{sec:physics:analytical}) that the residual mass discrepancy $\Delta_a$ (equation~\ref{eq:Delta_a-def}) is completely determined, with essentially \emph{no scatter}, by the ratio $\xi$ (equation~\ref{eq:xi-soln}) governing the quasi-adiabatic relaxation of dark matter in the presence of baryons in any galactic halo potential, through \eqn{eq:ardmByabary}. Since the physics of this relaxation can be approximated using simple fitting functions from the literature (equation~\ref{eq:X-def}), our framework allows us to analytically estimate both the median and scatter of the RAR ($\Delta_a$ as a function of \abary) in \emph{quantitative detail} over a wide dynamic range in galaxy and halo properties (Appendix~\ref{app:analytic}). 

We augmented our analytical calculations with measurements of the RAR in a realistic mock catalog of $\sim342,000$ low-redshift central galaxies with `baryonified' host halos produced using the algorithm of \citet[][PCS21]{pcs21} (sections~\ref{sec:mocks} and~\ref{sec:physics:mocks}).
We studied three regimes of \abary: (i) high-acceleration ($\abary\gtrsim10^{-10}\msq$), (ii) low-acceleration ($10^{-12}\msq\lesssim \abary\lesssim10^{-10}\msq$) and (iii) ultra-low-acceleration ($\abary\lesssim10^{-12}\msq$). Our main results can be summarized as follows.
\begin{itemize}
\item The \emph{median RAR} resulting from applying the relaxation prescription of \citet[][ST15]{st15} -- i.e., setting the relaxation parameter $q_{\rm rdm}=0.68$ in \eqn{eq:X-def} -- to our mock galaxies \emph{is within $\sim20\%$ of the observed relation} at low and high accelerations $\abary\gtrsim10^{-12}\msq$ (figure~\ref{fig:rar-fullsample}). 
Since the ST15 prescription and value of $q_{\rm rdm}$ were only tuned to reproduce the relaxation seen in halos in hydrodynamical CDM simulations, with no reference to the RAR, this quantitative agreement over more than four orders of magnitude in \abary\ represents a non-trivial success of the galaxy-dark matter association in $\Lambda$CDM.
\item This agreement is particularly remarkable in the high-acceleration regime, where we showed that  the median and scatter of the RAR are both very sensitive to the value of $q_{\rm rdm}$, and there is no \emph{a priori} reason why the value $q_{\rm rdm}=0.68$ should have worked (figure~\ref{fig:rar-fullsample-bkrxn} and Appendix~\ref{app:hernquist-rar}, see also below).
In this context, we also noted that there is presently some ambiguity in characterising the observed high-acceleration median RAR derived from rotation-supported and dispersion-supported galaxies \citep[e.g.][]{janz+16}, with the former being possibly lower than the latter \citep{cbsg19}. Adjusting $q_{\rm rdm}$ can track such differences (figure~\ref{fig:cubicRAR}), suggesting that high-acceleration RAR observations might place useful constraints on the physics of quasi-adiabatic relaxation, and hence on baryonic feedback prescriptions employed in cosmological hydrodynamical simulations (although see below).
\item The median RAR in the ultra-low-acceleration regime is very sensitive to the expelled (or diffuse) gas fraction $f_{\rm egas}$, and our default model predicts a distinctive break from smooth, MOND-inspired relations at $\abary\lesssim10^{-12}\msq$ for diffuse gas-rich systems (figure~\ref{fig:rar-diffusegas}). This regime, corresponding to the outskirts of halos hosting sub-$L_\ast$ galaxies, is currently unobserved, although future observations of the CGM could be promising in this regard \citep[e.g.,][]{cantalupo+14,werk+14,zahedy+19}. \emph{Our results at ultra-low-accelerations constitute robust and testable predictions of the $\Lambda$CDM framework.}
\item While the median RAR is set by a combination of baryon-dark matter scalings and relaxation physics (sections~\ref{subsec:relaxation} and~\ref{subsec:barycontent}, figures~\ref{fig:rar-coldgas} and~\ref{fig:rar-diffusegas}), we identified the primary source of \emph{scatter} in the RAR to be host halo mass and concentration, with a magnitude that depends on the value of the   relaxation parameter $q_{\rm rdm}$ (section~\ref{subsec:massconc} and Appendix~\ref{app:hernquist-rar}, figures~\ref{fig:rar-fullsample-bkrxn} and~\ref{fig:rar-massconc}). Specifically, the scatter in the high-acceleration, baryon-dominated regime is small when $q_{\rm rdm}\to1$ (perfect angular momentum conservation) and increases when $q_{\rm rdm}\to0$ (no baryonic backreaction).
So, in $\Lambda$CDM, the real puzzle posed by the observed tightness of the RAR is:  \emph{Why is $q_{\rm rdm}$ closer to 1 than to 0, with small scatter, over the relevant mass range?}
\item We used our mock galaxies to explore the sensitivity of the RAR to a number of details such as sample selection, rotation curve measurement technicalities, as well as variations in baryon-dark matter scalings and halo profile shape (section~\ref{subsec:details}). For example, we showed that the effect of a potential correlation between the stellar bulge size and halo concentration on the median RAR at high \abary\ can be degenerate with that of changing $q_{\rm rdm}$ (section~\ref{subsubsec:galscaling:stellar}), which must be kept in mind if the RAR is to constrain feedback physics as mentioned above. Our framework also provides a natural explanation for the observed offset \citep[][]{tian+20} between the RAR of cluster BCGs and fainter centrals (figure~\ref{fig:rar-luminosity} and section~\ref{subsubsec:lumcolour}), while predicting that the RAR is relatively stable against variations in the chosen form of the `un-baryonified' dark matter profile (NFW versus Einasto; section~\ref{subsubsec:dmprof}) or technicalities of rotation curve sampling (section~\ref{subsubsec:sampling}).
\item In contrast, we argued in section~\ref{sec:btfr} that the baryonic Tully-Fisher relation (BTFR) is substantially more susceptible to such technical details. As such, the RAR is a much more robust probe of galactic-scale gravitational physics than is the BTFR.
\end{itemize}

The intrinsic scatter of the RAR as inferred from observations, after accounting for all sources of measurement error, is a matter of considerable interest and discussion. If this scatter is indeed negligible, as reported by \citet{lmsp17}, it would pose a major challenge to the galaxy-dark matter association assumed in the $\Lambda$CDM paradigm. The robustness of this claim of a zero scatter RAR, however, remains debated \citep{rmdpd18,sc19,mrda20}. E.g., \citet{sc19} estimate an intrinsic scatter in the $\atot$-$\abary$ relation (they focus on the stellar contribution to \abary) of $0.11\pm0.02$ dex, fully consistent with our results above as well as those from the earlier $\Lambda$CDM literature \citep[e.g.,][]{kw17}. Uncertainties in observed rotation curves might also be sensitive to technical details of extracting velocity profiles \citep[e.g.,][]{sse21}. Finally, observational RAR (and BTFR) analyses often focus on `good' samples of inhomogeneously selected rotation curves \citep[e.g.,][]{lms16a,lmsp17}, making a direct comparison between predicted and observed  scatter difficult. 

Our results above therefore suggest that the median RAR (i.e., $\Delta_a$ as a function of \abary), especially in the regimes of high and ultra-low accelerations, is likely to be the most powerful discriminator between alternative gravitational models, as well as serving to constrain the physics of baryon-dark matter interactions in $\Lambda$CDM. For this to be successful, it will be important to perform observational analyses with well-defined, representative galaxy samples.


\section*{Acknowledgements}
AP thanks R. Srianand and Sowgat Muzahid for useful discussions.
We thank Kyu Chae for comments on an earlier draft, and our anonymous referee for an insightful report.
The research of AP is supported by the Associateship Scheme of ICTP, Trieste and the Ramanujan Fellowship awarded by the Department of Science and Technology, Government of India.
This work made extensive use of the open source computing packages NumPy \citep{vanderwalt-numpy},\footnote{\url{http://www.numpy.org}} SciPy \citep{scipy},\footnote{\url{http://www.scipy.org}} Matplotlib \citep{hunter07_matplotlib}\footnote{\url{https://matplotlib.org/}} and Jupyter Notebook.\footnote{\url{https://jupyter.org}}

\section*{Data Availability}
The mock catalogs underlying this work will be made available upon reasonable request to the authors.

\bibliography{references}

\appendix
\section{Analytic relaxation} \label{app:analytic}
Solving the relaxation problem boils down to describing how the final radius $r$ is related to the initial radius $r_{\rm in}$.  Equation~(\ref{eq:X-def}) of the main text considers a model in which 
\begin{equation}
    \frac{r}{r_{\rm in}} - 1 = q_{\rm rdm} \left(\frac{m_{\rm udm}(<r_{\rm in})}{m_{\rm tot}(<r)} - 1\right)
\end{equation}
where $m_{\rm udm}(<r_{\rm in})$ and $m_{\rm tot}(<r)$ are the initial and final enclosed mass profiles. Equation~\eqref{eq:mtot-def} says $m_{\rm tot}(<r) = m_{\rm bary}(<r) + f_{\rm rdm}\,m_{\rm udm}(<r_{\rm in})$ where $f_{\rm rdm}$ is the relaxed dark matter fraction (typically one sets $f_{\rm rdm} = 1-\Omega_{\rm b}/\Omega_{\rm m}$).  In the main text, this problem was treated numerically.  The main purpose of this Appendix is to show that, for judicious (but realistic) choices of the profile shapes $m_{\rm udm}(<r_{\rm in})$ and $m_{\rm bary}(<r)$, much of the analysis can be done analytically.  We assume spherical symmetry in what follows.

Previous analytic work \cite[e.g.][]{keeton2001} exploits the fact that, for simple choices of $m_{\rm bary}(<r)$, the relaxation equation can be solved analytically for any $m_{\rm udm}(<r_{\rm in})$.  This is attractive since, in practice, $m_{\rm udm}(<r_{\rm in})$ is unknown, so this is the quantity which one hopes to determine from detailed observations of the baryons.  In effect, such approaches solve for $r$ as a function of $m_{\rm udm}(<r_{\rm in})$, and hence of $r_{\rm in}$.  However, as we show below, for appreciating what sets the shape of the RAR, it is more illuminating to determine the inverse of this relation:  $r_{\rm in}$ as a function of $m_{\rm bary}(<r)$ and hence of $r$.  Below, we exploit the fact that, for simple choices of $m_{\rm udm}(<r_{\rm in})$ the post-relaxation profile can be written analytically for any $m_{\rm bary}(<r)$.  If, in addition, the $\abary(r) \equiv Gm_{\rm bary}(<r)/r^2$ vs $r$ relation can be inverted analytically, the result will be a fully analytic expression for the RAR from baryonic relaxation.

\subsection{The RAR for an initially Hernquist profile}
\label{app:hernquist-rar}
The main text used the NFW functional form to describe the initial profile, but also showed that an appropriately scaled Einasto profile gave very similar results.  Since the precise parametrization does not matter, it is natural to ask if there is a parametrization which simplifies the analysis.  For scales $r_{\rm in} < R_{\rm vir}$, the NFW model is very well approximated by a Hernquist profile, for which 
\begin{equation}
  \tilde m_{\rm udm}(r_{\rm in}) \equiv \frac{m_{\rm udm}(<r_{\rm in})}{m_{\rm vir}} = \left(\frac{r_{\rm in}}{R_{\rm vir}}\,\frac{R_{\rm vir} + r_1}{r_{\rm in} + r_1}\right)^2,
  \label{eq:hernquist}
\end{equation}
provided that one sets 
\begin{equation}
    c_1\equiv R_{\rm vir}/r_1 = c_{\rm NFW}^{0.75}/\sqrt{2}
\end{equation}
\citep{shds01}.  (Below, we will use the tilde to indicate mass profiles normalised by $m_{\rm vir}$.)

With equation~(\ref{eq:hernquist}) for $m_{\rm in}(r_{\rm in})$, the relaxation equation reads 
\begin{equation}
 \frac{q_{\rm rdm}}{\xi^3} = \left[1 - \frac{1-q_{\rm rdm}}{\xi}\right]\, \left[\frac{\tilde m_{\rm bary}(r)}{(1+1/c_1)^2}\left(\frac{1}{\xi} + \frac{R_{\rm vir}}{rc_1}\right)^2 + \frac{f_{\rm rdm}}{\xi^2}\right] ,
 \label{eq:cubicRAR}
\end{equation}
where $\xi\equiv r/r_{\rm in}$ as in the main text and we defined $\tilde m_{\rm bary}(r)\equiv m_{\rm  bary}(<r)/m_{\rm vir}$. This is a cubic equation for $\xi$ which can be solved analytically.  Since all the coefficients are real, there is at least one real root.  This root is given by 
\begin{equation}
    \frac{1}{\xi} = S - \frac{Q}{S} - a_2 ,
    \label{eq:root}
\end{equation}
where 
\begin{align}
    S &= \left[P + \sqrt{P^2 + Q^3}\right]^{1/3}, \nonumber\\
    P &= \frac{3 a_1 a_2 - a_0}{2} - a_2^3\quad {\rm and}\quad 
    Q = a_1 - a_2^2 ,\nonumber 
\end{align}
with 
\begin{align}
    a_2 &= \frac{2\mu\,(1-q_{\rm rdm})/(r/r_1) - (\mu + f_{\rm rdm})}{3a_3} ,\nonumber\\
    a_1 &= \frac{\mu}{r/r_1}\frac{(1-q_{\rm rdm})/(r/r_1) - 2}{3a_3} ,\nonumber\\
    a_0 &= -\frac{\mu}{(r/r_1)^2\,a_3}\quad {\rm and}\quad a_3 = q_{\rm rdm} + (1-q_{\rm rdm})(\mu + f_{\rm rdm}) ,\nonumber
\end{align}
and $\mu\equiv \tilde m_{\rm bary}(r)/(1 + 1/c_1)^2$.  
This follows from rearranging the cubic equation to $a_0+3a_1x+3a_2x^2+x^3=0$ and assumes $a_3\neq 0$ which is guaranteed if $q\leq1$.
Inserting this $\xi$ in equation~(\ref{eq:ardmByabary-thiswork}) yields the RAR for any input $\tilde m_{\rm bary}(r)$.  

\begin{figure}
\centering
\includegraphics[width=0.475\textwidth]{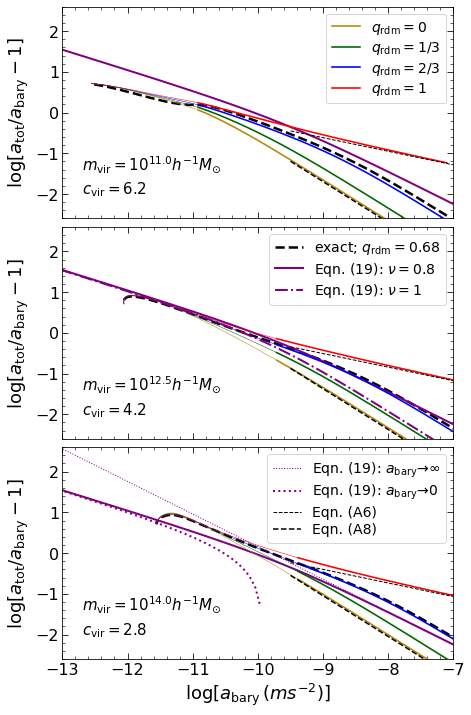}
\caption{Dependence of the RAR relations, determined from solving the cubic equation~(\ref{eq:cubicRAR}) with $m_{\rm bary}(r)$ from PCS21, on the relaxation parameter $q_{\rm rdm} = 0,1/3,2/3,1$ (solid coloured lines), for three choices of halo mass and associated halo concentration (\emph{top, middle} and \emph{bottom panels}; the corresponding mass profiles were shown in the left panel of figure~\ref{fig:relaxation-ratio}).
The thicker and thinner segments of these curves show scales smaller and larger, respectively, than that of the scale radius $h_{\Hi}$ of the \Hi\ gas. Thick dashed black curve in each panel shows the exact numerical solution using $q_{\rm rdm}=0.68$. Thick solid purple curve (repeated in each panel) shows \eqn{eq:RAR-simple} with $\nu=0.8$. 
Dotted purple curves in the \emph{bottom panel} show the associated scalings at large and small \abary.
Thick dot-dashed purple curve in the \emph{middle} panel shows \eqn{eq:RAR-simple} with $\nu=1$.
Thin dashed black curves show the predicted scalings for $q_{\rm rdm}=1$ and $q_{\rm rdm}=0$ (respectively, equation~\ref{eq:q=1} and~\ref{eq:q=0}) in the baryon-dominated regime.}
\label{fig:cubicRAR}
\end{figure}

Figure~\ref{fig:cubicRAR} shows the result if we use the $\tilde m_{\rm bary}(r)$ profiles (i.e. the sum of the stellar, \Hi, bound and expelled gas profiles) returned by the baryonification procedure of PCS21 for the three representative halo masses shown in figure~\ref{fig:relaxation-ratio}.  Note how the RAR changes as the quasi-adiabatic relaxation parameter varies (solid coloured curves) from $q_{\rm rdm}=0$ (no conservation) to $q_{\rm rdm}=1$ (exact conservation).  
For comparison, the thick dashed black curve in each panel shows the numerical solution for each (NFW-based) galaxy using $q_{\rm rdm}=0.68$; for the intermediate-mass halo, this is rather well-described by \eqn{eq:RAR-simple} with $\nu=0.8$. Interestingly, the $q_{\rm rdm}=1/3$ curve for the same object is equally well-described by \eqn{eq:RAR-simple} with $\nu=1$, which we discuss in the main text. For the two larger halo masses, the RAR turns over in the outer regions (small \abary).  This is because $\Delta_a = \tilde m_{\rm rdm}/\tilde m_{\rm bary}$ and figure~\ref{fig:relaxation-ratio} shows that $\tilde m_{\rm bary}$ increases more steeply than $\tilde m_{\rm rdm}$ in the outer regions of massive halos.  Note also that, at the smallest halo mass, the RAR flattens out, so that it lies below the scaling associated with $M_{\rm bary}\propto V_c^4$ for which $\atot\propto\abary^{1/2}$.  These results are in good {\em quantitative} agreement with those shown in the main text.  

Before moving on, it is worth noting that the structure of the cubic (equation~\ref{eq:cubicRAR}) makes it easy to understand how the RAR depends on $q_{\rm rdm}$, $f_{\rm rdm}$, $c_1$ and $m_{\rm vir}$ in the baryon-dominated limit.  For the PCS21 models with $q_{\rm rdm}>0$, this is where $r\to 0$ and $r\ll r_{\rm in}$.  In this $\xi\ll 1$ limit, the cubic becomes $\xi^{-3}\approx -a_0$. 
(Since $\atot/\abary \ge 1$ requires $\xi \ge 1-q_{\rm rdm}$, we treat the $q_{\rm rdm}\to 0$ limit more carefully below.)  
This is particularly instructive when $q_{\rm rdm}=1$, since then $a_3=1$ and $\xi = (1+c_1)^{2/3} [a_{\rm vir}/\abary(r)]^{1/3}$, where $a_{\rm vir}=Gm_{\rm vir}/R_{\rm vir}^2$. 
Moreover, we expect $f_{\rm rdm}\approx 1$, so equation~(\ref{eq:ardmByabary}) says that the RAR becomes 
\begin{equation}
  \Delta_a
  \approx 
  \frac{f_{\rm rdm}\, (1+c_1)^{2/3}}{[\abary(r)/a_{\rm vir}]^{1/3}} \quad{\rm when}\quad q_{\rm rdm}=1\,,
  \label{eq:q=1}
\end{equation}
where $\Delta_a$ was defined in \eqn{eq:Delta_a-def}.
This scaling is shown as the dashed line in figure~\ref{fig:cubicRAR}; clearly, it describes the approach to baryon domination well.  

We now consider $q_{\rm rdm}\to 0$ in the baryon-dominated regime, which is achieved by sending $r,r_{\rm in}\to0$ with $\xi\to1$.   For this, it is useful to rewrite equation~(\ref{eq:cubicRAR}) as 
\begin{equation}
 q_{\rm rdm} = (\xi - 1 + q_{\rm rdm})\, \left[\frac{m_{\rm bary}(r)}{(1+c_1)^2}\left(c_1 + \xi \frac{R_{\rm vir}}{r}\right)^2 + f_{\rm rdm}\right] ,
\end{equation}
making
\begin{align}
    \Delta_a 
    &= f_{\rm rdm}\,\left[\frac{m_{\rm bary}(r)}{(1+c_1)^2}\left(c_1 + \xi \frac{R_{\rm vir}}{r}\right)^2\right]^{-1}
    = f_{\rm rdm}\,\frac{m_{\rm udm}(r_{\rm in})}{m_{\rm bary}(r)}
    \label{eq:q=0}
    \\
    &\to f_{\rm rdm}\,\left[\frac{\abary(r)\,\xi^2}{a_{\rm vir}\,(1+c_1)^2}\right]^{-1} 
    \to \frac{f_{\rm rdm}\,(1+c_1)^2}{\abary(r)/a_{\rm vir}}\,\, {\rm when}\,\, q\to 0. 
    \nonumber
\end{align}
The second equality on the first line connects with equation~(\ref{eq:ardmByabary-pert}) of the main text, and the final expression is from the $r_{\rm in}\to 0$ and $\xi\to 1$ limits.  This has a different dependence on \abary\ than when $q_{\rm rdm}=1$, but the dashed curve in figure~\ref{fig:cubicRAR} (which is almost indistinguishable from the magenta part of the $q_{\rm rdm}=0$ curve) shows that it describes this limit well.  

Notice that the quantity $a_{\rm vir}\,(1 + c_1)^2$ plays a key role.  Since all halos have density $200\rho_{\rm crit}$, 
\begin{equation}
 \frac{a_{\rm vir}}{\rm nm\, s^{-2}} = 0.0032\,\frac{R_{\rm vir}}{100h^{-1}{\rm kpc}}
  = 0.0053\, \left(\frac{m_{\rm vir}}{10^{12} h^{-1}M_\odot}\right)^{1/3}.
 \label{eq:achar}
\end{equation}
For $m_{\rm vir}=10^{12}h^{-1}M_\odot$ and $c_{\rm vir}=10$, typical of halos hosting galaxies, 
$a_{\rm vir}\,(1 + c_1)^2\approx 0.13$~nm~s$^{-2}$.  Since $f_{\rm rdm}\approx 1$, this explains the $a_0=0.12$~nm~s$^{-2}$ scale that is usually associated with the RAR from galactic dynamics. Of course, equation~(\ref{eq:achar}) shows that this scale will be larger for clusters.  Thus, our analysis shows that the RAR {\em scale} depends on halo mass, concentration and dark matter fraction, but the {\em shape} of the RAR in the high-acceleration regime depends on the adiabatic parameter $q_{\rm rdm}$.  

The analysis above is also useful for understanding why the RAR has small scatter.  At fixed $f_{\rm rdm}$ (and $q_{\rm rdm}$) the scatter comes from $m_{\rm vir}^{1/3}\,(1 + c_1)^2$.  However, in $\Lambda$CDM, more massive halos are less concentrated, so averaging over a factor of $\sim 10$ in $m_{\rm vir}$ does not lead to large scatter in the RAR.  (Of course, variations of order $10^3$ in $m_{\rm vir}$ will be more significant, which is why the RAR of clusters is offset from that of galaxies -- but by much less than a factor of $10^3$.)  Moreover, equations~(\ref{eq:q=1}) and~(\ref{eq:q=0}) depend on different powers of this combination of mass and concentration; this explains why the scatter around the $q_{\rm rdm}=0$ relation is larger than around $q_{\rm rdm}=1$ (c.f. figure~\ref{fig:rar-fullsample-bkrxn}).  This leaves variations in $f_{\rm rdm}$ and $q_{\rm rdm}$ as possible additional sources of scatter in the RAR.  However, $f_{\rm rdm}$ is expected to have small scatter -- and in our mocks it has {\em no} scatter (by assumption). So, in $\Lambda$CDM, the real puzzle posed by the tightness of the RAR is:  Why is $q_{\rm rdm}$ closer to 1 than to 0, with small scatter, over the mass range relevant to the RAR of galaxies?

\subsection{Fully analytic relaxation and the RAR}
Although we used $m_{\rm bary}(<r)$ from PCS21 to make figure~\ref{fig:cubicRAR}, the analysis of the previous subsection applies to any $m_{\rm bary}$.  This means that, for judicious parameterizations of $m_{\rm bary}$, it may be possible to provide fully analytic expressions for the RAR.  We now show that this is indeed possible over a substantial fraction of the halo.  

\begin{figure}
\centering
\includegraphics[width=0.475\textwidth]{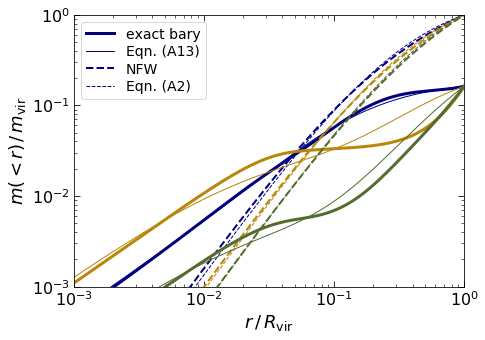}
\caption{Comparison of the actual profiles (thick curves) with the analytic approximations (thin curves) described in the text for the three halos considered in figure~\ref{fig:cubicRAR}, with the solid (dashed) curves showing baryonic (initial dark matter) profiles. The colour-coding is identical to that in figure~\ref{fig:relaxation-ratio}. For all halos, \eqn{eq:hernquist} describes the initial profiles well.  Equation~(\ref{eq:mbar}) describes $m_{\rm bary}$ of the least massive halo (blue solid curves) well out to a substantial fraction of the virial radius, but the gas in the more massive halos is less centrally concentrated, so the agreement is only good out to the scale radius of the gas (c.f. figure~\ref{fig:cubicRAR}).}
\label{fig:mprofile}
\end{figure}

Start with profiles of the form 
\begin{equation}
  \rho_\beta(r) \propto \frac{(r/r_\beta)^{-\beta}}{(1 + r/r_\beta)^{4-\beta}},
\end{equation}
which scale as $r^{-\beta}$ on scales smaller than $r_\beta$, and as $r^{-4}$ on larger scales.  Simulations have shown that it is reasonable to approximate the stellar distribution with $\beta=2$, the \Hi\ gas with $\beta=1$ and the bound and expelled gas profiles with $\beta=0$ (but different scale radii $r_\beta$).  
Let 
\begin{equation}
    M_\beta \equiv 4\pi \int_0^\infty dx\, x^2\,\rho_\beta(x) 
\end{equation}
denote the total mass associated with this profile.  
Then, provided $\beta < 3$, the mass within $r$ is given by 
\begin{equation}
  M_\beta(<r) = M_\beta\,\left(\frac{r/r_\beta}{1 + r/r_\beta}\right)^{3-\beta}.
\end{equation}
If we ignore the two $\beta=0$ components -- in practice we assign their mass to the $\beta=1$ component and modify $r_1$ to match the profile of their sum -- then 
\begin{equation}
    \tilde m_{\rm bary}(r) \equiv \frac{M_{\rm bary}(r)}{m_{\rm vir}} = F_2\, \frac{r}{r + r_2} + F_1\,\frac{r^2}{(r + r_1)^2},
    \label{eq:mbar}
\end{equation}
where $F_\beta\equiv M_\beta/M_{\rm tot}$.  It is conventional to work not with $F_\beta$ but with the mass fractions within the virial radius:
\begin{equation}
   f_\beta \equiv \frac{M_\beta(<R_{\rm vir})}{m_{\rm vir}} = F_\beta\, \left(\frac{R_{\rm vir}/r_\beta}{1 + R_{\rm vir}/r_\beta}\right)^{3-\beta}.  
\end{equation}
Then 
\begin{equation}
 F_\beta = f_\beta\, (1 + r_\beta/R_{\rm vir})^{3-\beta} \quad {\rm and}\quad 
 f_d\equiv 1 - \sum_\beta f_\beta .
\end{equation}
Equation~(\ref{eq:mbar}) is substantially more realistic, and not much more complicated, than the model discussed in Appendix A of \citet{teyssier+11} in which all the baryons are clubbed into a single component with $\rho\propto r^{-1}$.  Figure~\ref{fig:mprofile} illustrates.  Thick and thin solid curves compare the actual and approximated $\tilde m_{\rm bary}$ profiles for the two  halos shown in figure~\ref{fig:cubicRAR}.  To produce the curves, we set $r_1=h_{\Hi}$ (the \Hi\ disk scale length) for the lower mass halo and $r_1=r_{\rm s}$ (the NFW scale radius) for the higher mass halo. For the least massive halo, the agreement is good over almost the entire halo, whereas, for the more massive halos, the agreement is good only in the inner regions which are dominated by stars, and out to about the scale radius of the gas.  Thick and thin dashed curves show that equation~(\ref{eq:hernquist}) provides a good description of the initial NFW profile for $\tilde m_{\rm udm}$.  
 
If the baryonic profile is well described by equation~(\ref{eq:mbar}) then  
\begin{equation}
 \frac{\abary(r)}{a_{\rm vir}} = F_2\, \frac{R_{\rm vir}^2}{r(r + r_2)} + F_1\,\frac{R_{\rm vir}^2}{(r + r_1)^2}.
 \label{eq:abar}
\end{equation}
Equation~(\ref{eq:abar}) shows that, to invert the $\abary$-$r$ relation one must solve a quartic equation.  Therefore, $r$ can be written as a complicated but analytic function of $\abary$, which, when inserted for $r$ in equation~(\ref{eq:root}) yields $\xi$ as a function of $\abary$.  This $\xi(\abary)$, when inserted in equation~(\ref{eq:ardmByabary-thiswork}), yields a fully analytic expression for the RAR.  The accuracy of this expression depends on how well equation~(\ref{eq:mbar}) approximates the true $m_{\rm bary}(r)$.  Figure~\ref{fig:mprofile} shows that we expect this to work well out to approximately the scale where the gas dominates the baryonic component.  As this fully analytic RAR is essentially indistinguishable from that shown by the magenta parts of the curves in figure~\ref{fig:cubicRAR}, we have not shown it again.

The procedure just described yields an analytic RAR by first finding $r_{\rm in}$ as a function of $r$, and then writing $r$ as a function of $\abary$.  For $\tilde m_{\rm bary}(r)$ given by equation~(\ref{eq:mbar}), it is also possible to do the opposite.  I.e., equations~(\ref{eq:xi-soln}) and~(\ref{eq:X-def}) yield 
a quartic equation for $r$, which can be solved analytically to yield $r/r_{\rm in}$ for any $m_{\rm udm}(<r_{\rm in})$.  
This solution for $r$ can be inserted in equation~(\ref{eq:abar}) to yield $\abary$ and then 
\begin{align}
 \Delta_a\equiv \frac{\atot}{\abary} - 1 &= \frac{f_{\rm rdm} \tilde m_{\rm udm}(r_{\rm in})}{m_2(r) + m_1(r)} \\
  &= \frac{f_{\rm rdm} \tilde m_{\rm udm}(r_{\rm in})\, (1+r_2/r)(1+r_1/r)^2}{F_2\, (1 + r_1/r)^2 + F_1\, (1 + r_2/r)} \nonumber
\end{align}
yields the corresponding fully analytic (but messy!) RAR.  Over the range where equation~(\ref{eq:mbar}) provides a good description of $\tilde m_{\rm bary}(r)$, the $r_{\rm in}(r)$ and $r(r_{\rm in})$ approaches are almost indistinguishable.  However, the $r_{\rm in}(r)$ approach, in which $\tilde m_{\rm udm}$ is given by equation~(\ref{eq:hernquist}) and $\tilde m_{\rm bary}$ is arbitrary, is more efficient (solve a cubic rather than quartic).  

We conclude that we have analytic understanding of all the RAR scalings presented in the main text.

\label{lastpage}

\end{document}